\documentclass[sigconf]{acmart}

\AtBeginDocument{
  \providecommand\BibTeX{{
    \normalfont B\kern-0.5em{\scshape i\kern-0.25em b}\kern-0.8em\TeX}}}

\setcopyright{acmcopyright}
\copyrightyear{2018}
\acmYear{2018}
\acmDOI{10.1145/1122445.1122456}

\acmConference[ACSAC'20]{ACSAC'20: Annual Computer Security Applications Conference}{December 07--11, 2020}{Austin, TX}
\acmBooktitle{ACSAC'20: Annual Computer Security Applications Conference, December 07--11, 2020 Austin, TX}
\acmPrice{15.00}
\acmISBN{978-1-4503-XXXX-X/18/06}

\usepackage{xspace}
\newcommand{\ie}{i.\@\,e.,\@\xspace}
\newcommand{\eg}{e.\@\,g.,\@\xspace}

\newcommand{\etal}{et~al.\@\xspace}

\usepackage{amsmath}
\usepackage[noend]{algpseudocode}
\usepackage{algorithm}

\usepackage{epigraph}

\usepackage{booktabs}
\usepackage{multirow}

\newcommand{\new}[1]{{#1}}
\newcommand{\old}[1]{}

\newcommand{\oldtext}[1]{}

\newcommand{\camera}[1]{{#1}}

\copyrightyear{2020} 
\acmYear{2020} 
\setcopyright{licensedothergov}\acmConference[ACSAC 2020]{Annual Computer Security Applications Conference}{December 7--11, 2020}{Austin, USA}
\acmBooktitle{Annual Computer Security Applications Conference (ACSAC 2020), December 7--11, 2020, Austin, USA}
\acmPrice{15.00}
\acmDOI{10.1145/3427228.3427276}
\acmISBN{978-1-4503-8858-0/20/12}

\begin{document}

\title{\textsc{Imperio}: Robust Over-the-Air Adversarial Examples for Automatic Speech Recognition Systems}

\author{Lea Sch\"onherr, Thorsten Eisenhofer, Steffen Zeiler, Thorsten Holz, and Dorothea Kolossa}
\affiliation{Ruhr University Bochum}
\email{{lea.schoenherr,thorsten.eisenhofer,steffen.zeiler,thorsten.holz,dorothea.kolossa}@rub.de}

\begin{abstract}
Automatic speech recognition (ASR) systems can be fooled via targeted adversarial examples, which induce the ASR to produce arbitrary transcriptions in response to altered audio signals. However, state-of-the-art adversarial examples typically have to be fed into the ASR system directly, and are not successful when played in a room. Previously published over-the-air adversarial examples fall into one of three categories: they are either handcrafted examples, they are so conspicuous that human listeners can easily recognize the target transcription once they are alerted to its content, or they require precise information about the room where the attack takes place, and are hence not transferable to other rooms.

In this paper, we demonstrate the first algorithm that produces \emph{generic} adversarial examples \new{against hybrid ASR systems}, which remain \emph{robust} in an over-the-air attack that is \emph{not adapted} to the specific environment. Hence, no prior knowledge of the room characteristics is required. Instead, we use \emph{room impulse responses} (RIRs) to compute robust adversarial examples for arbitrary room characteristics and employ the ASR system \emph{Kaldi} to demonstrate the attack. Further, our algorithm can utilize psychoacoustic methods to hide changes of the original audio signal below the human thresholds of hearing. In practical experiments, we show that the adversarial examples work for varying room setups, and that no direct line-of-sight between speaker and microphone is necessary. As a result, an attacker can create \emph{inconspicuous} adversarial examples for \emph{any} target transcription and apply these to \emph{arbitrary} room setups without any prior knowledge.
\end{abstract}


\maketitle

\section{Introduction}
\epigraph{In restless dreams I walked alone. Narrow streets of cobblestone. 'Neath the halo of a streetlamp. I turned my collar to the cold and damp. When my eyes were stabbed by the flash of a neon light. That split the night. And touched the sound of silence.}{Simon \& Garfunkel, \emph{The Sound of Silence}}

% WHAT IS THE PROBLEM DOMAIN?
Substantial improvements in speech recognition accuracy have been achieved in recent years by using acoustic models based on deep neural networks (DNNs). Nevertheless, current studies suggest that there can be significant differences in the mechanism of artificial neural network algorithms compared to human expectations. This is a very unfortunate situation, as a rogue party can abuse this knowledge to create input data, which leads to inconsistent recognition results, without being noticed~\cite{carlini2017towards, carlini2018audio}. As just one example of such attacks, several recent works have demonstrated that it is possible to fool different kinds of ASR systems into outputting a malicious transcription chosen by the attacker~\cite{carlini2018audio,yuan-18-commandersong,Schoenherr2019,qin-19-robust,carlini2016hidden, Abdullah-19-practical,yakura-2019-robust, szurley-2019-perceptual}.

\begin{figure}
    \centering
  \includegraphics[width=1.0\columnwidth]{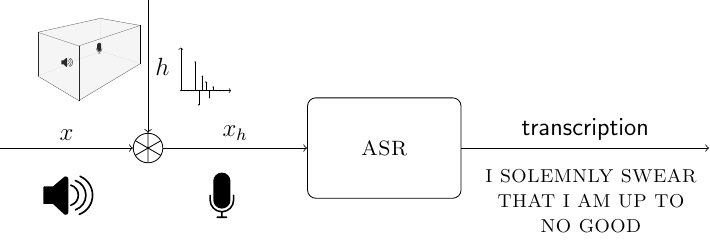}
    \caption{For an over-the-air attack against automatic speech recognition (ASR) systems, the attack should remain viable after the transmission over the air. This transmission can be modeled as a convolution of the original audio signal~$x$ with the \emph{room impulse  response} (RIR)~$h$.}
    \label{fig:overview}
\end{figure}

The practical implications and real-world impact of the demonstrated attacks are unclear at the moment. On the one hand, earlier work fed the adversarial audio examples \emph{directly} into the ASR system~\cite{carlini2018audio,Schoenherr2019,yuan-18-commandersong}, hence ignoring all side effects (e.\,g., echo or reverberation) of a real-world environment, where the sound is transmitted from a loudspeaker to the input microphone of the recognition engine. On the other hand, some works demonstrated adversarial examples that can be played over-the-air~\cite{carlini2016hidden, Abdullah-19-practical, szurley-2019-perceptual, yakura-2019-robust}, but these proof-of-concept attacks are either tailored to a single, static room setup or are hard to reproduce systematically with a proven success rate in a different environment like the attack sketched in CommanderSong~\cite{yuan-18-commandersong}. Recently and independently, Chen~\etal~\cite{chen-2020-metamorph} showed a first over-the-air attack. Their attack was evaluated against \emph{DeepSpeech}~\cite{hannun-14-deepspeech}. In contrast, we are showing an attack against Kaldi, a hybrid ASR system, based on a combination of a DNN-based acoustic model and a subsequent search for the optimal word sequence in a weighted-finite-state-transducer model. This approach is conceptually completely different from the end-to-end approach in \emph{DeepSpeech}, as used by Chen~\etal Hybrid systems such as Kaldi are significant here as they show the best performance on many speech recognition tasks, require comparatively little training material, allow for an easy replacement of task-specific grammars, and are therefore widely adopted in the
~industry.
 
In cases where over-the-air adversarial examples have been used in black-box settings, the target transcription is easily perceived by human listeners, once the intended attack is known~\cite{carlini2016hidden, Abdullah-19-practical}. 

% WHAT MAKES THE PROBLEM INTERESTING / HARD?
We argue that adversarial examples for ASR systems can only be considered a real threat if the targeted recognition is produced even when the signal is played over the air. 
Compared to previous attacks, where the manipulated speech signal is fed directly into the ASR system, over-the-air attacks are more challenging, as the transmission over the air significantly alters the signal. 

Our key insight that forms the basis of this paper is that this transmission can be modeled as a convolution of the original audio signal with the \emph{room impulse response}~(RIR), which describes the alterations of an acoustic signal by the transmission via loudspeaker to the microphone (see Figure~\ref{fig:overview} for an illustration), where the RIR depends on various factors~\cite{allen-1979-image}. In practice, it is nearly impossible to estimate an exact RIR without having access to the actual room. Therefore, robust adversarial examples need to take a range of possible RIRs into account to increase the success rate. Nevertheless, we show that for a successful attack, it is not necessary to acquire precise knowledge about the attack setup; instead, a generic adversarial example computed for a large variety of possible rooms is
~enough.

% WHAT IS THE STATE-OF-THE-ART IN THIS DOMAIN?
\textbf{Robust Adversarial Examples.}
The first adversarial audio examples imperceptible to humans, even if they know the target transcription, have been described by Carlini and Wagner~\cite{carlini2018audio}. Other approaches~\cite{Schoenherr2019,qin-19-robust} have been successful at embedding most changes below the human threshold of hearing, which makes them much harder to notice. On the downside, none of these attacks were successfully demonstrated when played over the air as the adversarial examples need to be fed directly into the ASR system.

Approaches, which did work over the air, have only been tested in a static setup (\ie fixed position of speaker and microphone with a fixed distance). 
Yakura's and Sakuma's~\cite{yakura-2019-robust} approach can hide the target transcription but requires physical access to the room to playback the audio while optimizing the adversarial example, which limits their attack to one very specific room setup and is very time costly. Szurley and Kolter~\cite{szurley-2019-perceptual} published room-dependent robust adversarial examples, which even worked under constraints given by a psychoacoustic model, describing the human perception of sound. However, their adversarial examples have only been evaluated in an anechoic chamber (i.\,e., a room specifically designed to absorb reflections). The attack can, therefore, not be used in real-world scenarios, but only in carefully constructed laboratories with properties that are never given in natural environments.
In other successful over-the-air attacks, human listeners can easily recognize the target transcription once they are alerted to its content~\cite{carlini2016hidden, Abdullah-19-practical}.  Chen~\etal~\cite{chen-2020-metamorph} showed a first over-the-air attack against the end-to-end recognition system \emph{DeepSpeech}~\cite{hannun-14-deepspeech}, relying on a database of measured room transfer functions.

In contrast, our approach is inspired by Athalye et al.'s seminal work: 
A real-world 3D-printed turtle, which is recognized as a rifle from almost every point of view due to an adversarial perturbation~\cite{athalye2017synthesizing}. The algorithm for creating this 3D object not only minimizes the distortion for one image, but for \emph{all} possible projections of a 3D object into a 2D image. We borrow the idea and transfer it to the audio domain, replacing the projections by convolutions with RIRs, thereby hardening the audio adversarial example against the transmission through varying rooms.

% WHAT IS OUR APPROACH / CONTRIBUTION?
\textbf{Contributions.}
With \textsc{Imperio}, we introduce the first method to compute generic and robust over-the-air adversarial examples against hybrid ASR systems . We achieve this by utilizing an RIR generator to sample from different room setups. We implement a full, end-to-end attack that works in both cases, with and without psychoacoustic hiding. In either case, we can produce successful robust adversarial examples. With our generic approach, it is possible to induce an arbitrary target transcription in any kind of audio without physical access to the target room.

More specifically, for the simulation, the convolution with the sampled RIR is added as an additional layer to the ASR's underlying neural network, which enables us to update the original audio signal directly under the constraints given by the simulated RIR. For this purpose, the RIRs are drawn out of a distribution of room setups to simulate the over-the-air attack. Using this approach, adversarial examples are hardened to remain robust in real over-the-air attacks across various room setups. We also show a reduction of the added perturbations based on \emph{psychoacoustic hiding}~\cite{zwicker2013psychoacoustics}, by including hearing thresholds in the backpropagation, as proposed by Sch\"onherr et al.~\cite{Schoenherr2019}.

% HIGH-LEVEL OVERVIEW OF RESULTS
We have implemented the proposed algorithm to attack the hybrid DNN-HMM ASR system \emph{Kaldi}~\cite{Povey_ASRU2011_2011} under varying room conditions. We demonstrate that generic adversarial examples can be computed that are transferable to different rooms and work without line-of-sight, distances in the range of meters, and even if the microphone records no direct sound but only a reflection. In fact, we even show that our generic approach, using only simulated RIRs, creates more robust adversarial examples compared to real measured examples indicating that no prior knowledge about the attack setup is required for our attack.

\begin{figure*}
    \centering
   \includegraphics[width=1.8\columnwidth]{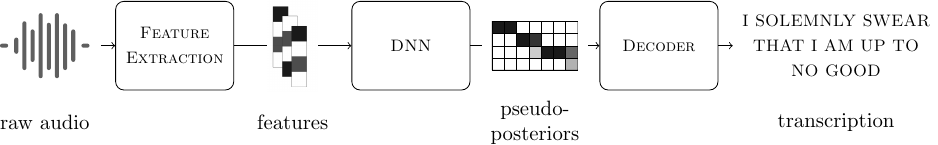}
    \caption{Overview of a state-of-the-art hybrid ASR system with the three main components of the ASR system: feature extraction, calculating pseudo-posteriors with a DNN, and decoding.}
    \label{fig:asr_overview}
\end{figure*}

% CONTRIBUTIONS
\smallskip \noindent
In summary, we make the following three key contributions:

\begin{itemize}
    \item \textbf{Robust Over-The-Air Attack.} We propose a generic approach to generate robust over-the-air adversarial examples for DNN-HMM-based ASR systems. The attack uses a DNN convolution layer to simulate the effect of arbitrary RIRs, which allows us to alter the raw audio signal directly. 
    \item \textbf{Psychoacoustics.} We show that the attack can be combined with psychoacoustic methods for reducing the perceived distortions.
    \item \textbf{Performance Analysis.} We evaluate the success rate of the adversarial attack and analyze the amount of added perturbation. We investigate the influence of increasing reverberation time, increasing microphone-to-speaker distances, different rooms, and no direct line-of-sight between speaker and microphone.
\end{itemize}

A demonstration of our \textsc{Imperio} attack is available online at \camera{\url{http://imperio.adversarial-attacks.net}}, where we present several adversarial audio files which have been successfully tested when played over-the-air.

\section{Background}

In the following, we provide an overview of the ASR system that we used in the attack and describe the general approach to calculate audio adversarial examples. Furthermore, we discuss how room simulations can be performed with the help of RIRs and briefly introduce the necessary background from psychoacoustics as these are used to hide the attack.

\subsection{Automatic Speech Recognition}

For the demonstration of an end-to-end attack, we chose the open-source speech recognition toolkit \emph{Kaldi}~\cite{Povey_ASRU2011_2011}, which has been used in previous attacks~\cite{yuan-18-commandersong, Schoenherr2019} and is also used in commercial tools like Amazon's Alexa~\cite{Schoenherr2019}. 
In Figure~\ref{fig:asr_overview}, a high-level overview of this system is given. The DNN-HMM-based ASR system can be divided into three parts: the feature extraction, which transforms the raw input data into representative features, the DNN as the acoustic model of the system, and the decoding step, which returns the recognized transcription. 

\begin{figure}[t]
    \centering
  \includegraphics[width=1.0\columnwidth]{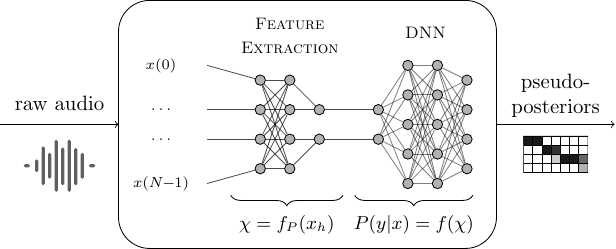}
    \caption{Augmented DNN, which gets the raw audio as input and integrates the feature extraction into the recognizer's DNN. This enables us to update the raw audio signal directly via gradient descent.}
    \label{fig:dnn}
\end{figure}

\paragraph{Feature Extraction} For the feature extraction, the raw audio is divided into frames (\eg $20$\,ms long) with a certain overlap (\eg $10$\,ms) between two neighboured frames. For each of these frames, a discrete Fourier transform (DFT) is performed to retrieve a frequency representation of the audio input. Next, the magnitude and the logarithm of the resulting complex signal are calculated. The result is a common representation of audio features in the frequency domain. 
In  Sch\"onherr et al.'s approach, this feature extraction is integrated into the DNN, allowing them to directly modify the raw audio data when computing adversarial examples (see Figure~\ref{fig:dnn} for an illustration). 

\paragraph{Acoustic Model DNN} The features described above are used as the input for the acoustic model DNN. Based on these, the DNN calculates a matrix of so-called pseudo-posteriors, which describe the probabilities for each of the phones of the language---English, in this case---being present in each time step $t = 1 \dots T$. 

\paragraph{Decoding} Finally, the pseudo-posteriors are used to calculate the most likely transcription via Viterbi decoding and an HMM-based language model.

\smallskip
This so-called \emph{hybrid} approach, which realizes speech recognition through a search for the most likely path through a matrix of phone posteriors, is easier to train and still achieves better results in comparison to end-to-end approaches~\cite{luscher-19-rwth}.

\subsection{Adversarial Audio Examples}
For the calculation of adversarial examples, the ASR system can be described as the function 
\begin{equation}
y = \arg \max_{\tilde{y}}P(\tilde{y}|x) = f(x),
\end{equation}
mapping an audio signal $x$ to its corresponding, most likely transcription~$y$. An adversarial example is generated by modifying the original input
\begin{equation}
x' = x + \delta, \quad \text{such that} \quad f(x) \neq f(x').
\end{equation}
The added distortions~$\delta$ can also be restricted, \eg via hearing thresholds. In this work, only targeted attacks are considered, where the target transcription~$y'\stackrel{!}{=}f(x')$ is defined by the attacker. The optimization can, therefore, be described as
\begin{equation}
x' = \arg \max_{\tilde{x}} P(y'|\tilde{x}).
\label{eq:optimization}
\end{equation}
To calculate robust over-the-air adversarial examples, we base our work on the approach proposed by Sch\"onherr \etal~\cite{Schoenherr2019} and similar works. The method can be divided into three steps: forced alignment, gradient descent, and restriction of the perturbations via hearing thresholds. 

\paragraph{Forced Alignment} Forced alignment is typically used during the training of the ASR systems when no exact alignments between the audio and the transcription data are available. In our case, we utilize this algorithm to find the best possible alignment of the original audio input and the malicious target transcription. 

\paragraph{Gradient Descent} For the attack, the feature extraction is integrated into the DNN, so that the raw audio data can be updated directly via gradient descent. For this purpose, the cross-entropy loss is measured between the target output---the pseudo-posteriors---and the actual DNN output, and is used to compute gradients for the optimization algorithm.

\paragraph{Psychoacoustic Hearing Thresholds} The added noise is restricted to those time-frequency ranges, where noise perceptibility is minimal. For this, we use psychoacoustic hearing thresholds described by Zwicker and Fastl~\cite{zwicker2013psychoacoustics}.

\subsection{Room Impulse Response}

When an audio signal is transmitted through a room, as visualized in Figures~\ref{fig:overview} and \ref{fig:room}, the recorded signal can be approximated by convolving the room's impulse response $h$  with the original audio signal $x$ as
\begin{equation}
x_h = x \ast h.
\end{equation}
Here,  the convolution operator $\ast$ is a shorthand notation for the multi-path transmission model

\begin{align}\label{eq:defconv}
\begin{split}
x_h(n) = &\sum_{m = n-M+1}^{n} x(m) \cdot h(n-m)\\
&\text{with} \quad n = 0, \dots, N-1,
\end{split}
\end{align}
where $N$ is the length of the audio signal, $M$ the length of the RIR~$h$, and all $x(n)$ with $n < 0$ are assumed to be zero.

 In general, the RIR~$h$ depends on the size of the room, the positions of the source and the receiver, and other room characteristics such as the sound reflection properties of the walls, any furniture, people, or other contents of the room. Hence, the audio signal received by the ASR system is never identical to the original audio, and an exact RIR is practically impossible to predict. We describe a possible solution for a sufficient approximation in Section~\ref{sec:approach}.

\subsection{Psychoacoustics}

Psychoacoustics yields an effective measure of \mbox{(in-)audibility}, which is also helpful for the calculation of inconspicuous audio adversarial examples~\cite{Schoenherr2019, qin-19-robust}. 
Psychoacoustic hearing thresholds describe how the dependencies between frequencies and across time lead to masking effects in human perception~\cite{zwicker2013psychoacoustics}. Probably the best-known example for an application of these effects is found in MP3 compression~\cite{ISO11172}, where the compression algorithm uses empirical hearing thresholds to minimize bandwidth or storage requirements. For this purpose, the original input signal is transformed into a smaller but lossy representation.  

For an attack, the psychoacoustic hearing thresholds are used to limit the changes in the audio signal to time-frequency-ranges, where the added perturbations are not, or barely, perceptible by humans. 
To calculate the hearing thresholds, we use the approach described by Sch\"onherr \etal~\cite{Schoenherr2019}.
\section{Over-the-Air Adversarial Examples}
\label{sec:approach}

Our goal is to compute robust audio adversarial examples, which still work after transmission from a loudspeaker. For this,  we simulate different RIRs and employ an iterative algorithm to compute adversarial examples robust against signal modifications that are incurred during playback in a room.
 
\subsection{Threat Model}
\label{sec:attack}
Throughout the rest of this paper, we consider the following threat model similar to prior work in this area. We assume a white-box attack, where the adversary knows the internals of the ASR system, including all its model parameters. This requirement is in line with prior work on this topic~\cite{Schoenherr2019, carlini2018audio, yuan-18-commandersong}. Using this knowledge, the attacker generates malicious audio samples offline before the actual attack takes place, \ie the attacker exploits the ASR system to create an audio file that produces the desired recognition result, which is then played via a loudspeaker. Additionally, we assume that the trained ASR system, including the DNN, remains unchanged over time. Finally, we assume that the adversarial examples are played over the air. Note that we only consider targeted attacks, where the target transcription is predefined (i.e., the adversary chooses the target transcription). Finally, we assume a threat model where a potential attacker can run an extensive search. Specifically, the attacker is able to calculate a batch of potential adversarial examples and select those examples that are especially robust.

\subsection{Room Impulse Response Simulator}
\label{rirsimulator}

\begin{figure}
  \centering
  \includegraphics[width=1.0\columnwidth]{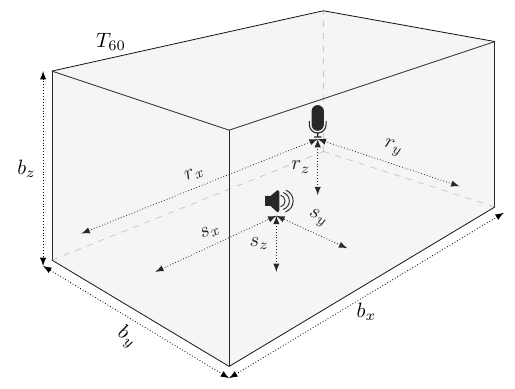}
  \caption{For the room simulation model, we assume a probability distribution over all possible rooms by defining relevant simulation parameters like the room geometry, the reverberation time~$T_{60}$, and positions of source and receiver as random variables. To optimize our over-the-air adversarial examples, we sample from this distribution to get a variety of possible RIRs.}
  \label{fig:room}
\end{figure}

To simulate RIRs, we use the \emph{AudioLabs} implementation based on the image method from Allen and Berkley~\cite{allen-1979-image}. The simulator takes as input the room dimensions, the reverberation time~$T_{60}$, and the position of source and receiver and approximates the corresponding RIR for the given parameters.

For our attack, we model cuboid-shaped rooms, which can be described by their length~$b_x$, width~$b_y$, height~$b_z$ defined as $\mathbf{b}=[b_x, b_y, b_z]$. In addition to this, we model the three-dimensional source position~$\mathbf{s}=[s_x, s_y, s_z]$, receiver position~$\mathbf{r}=[r_x, r_y, r_z]$, and the reverberation time~$T_{60}$, which is a standard measure for the audio decay time, defined as the time it takes for the sound pressure level to reduce by $60$\,dB. This results in ten freely selectable parameters. All parameters are also sketched in Figure~\ref{fig:room}. Even though this might seem like an overly simple model, we show that the computed adversarial examples are indeed robust for real rooms that are more
~complex.

\begin{figure}
    \centering
  \includegraphics[width=1.0\columnwidth]{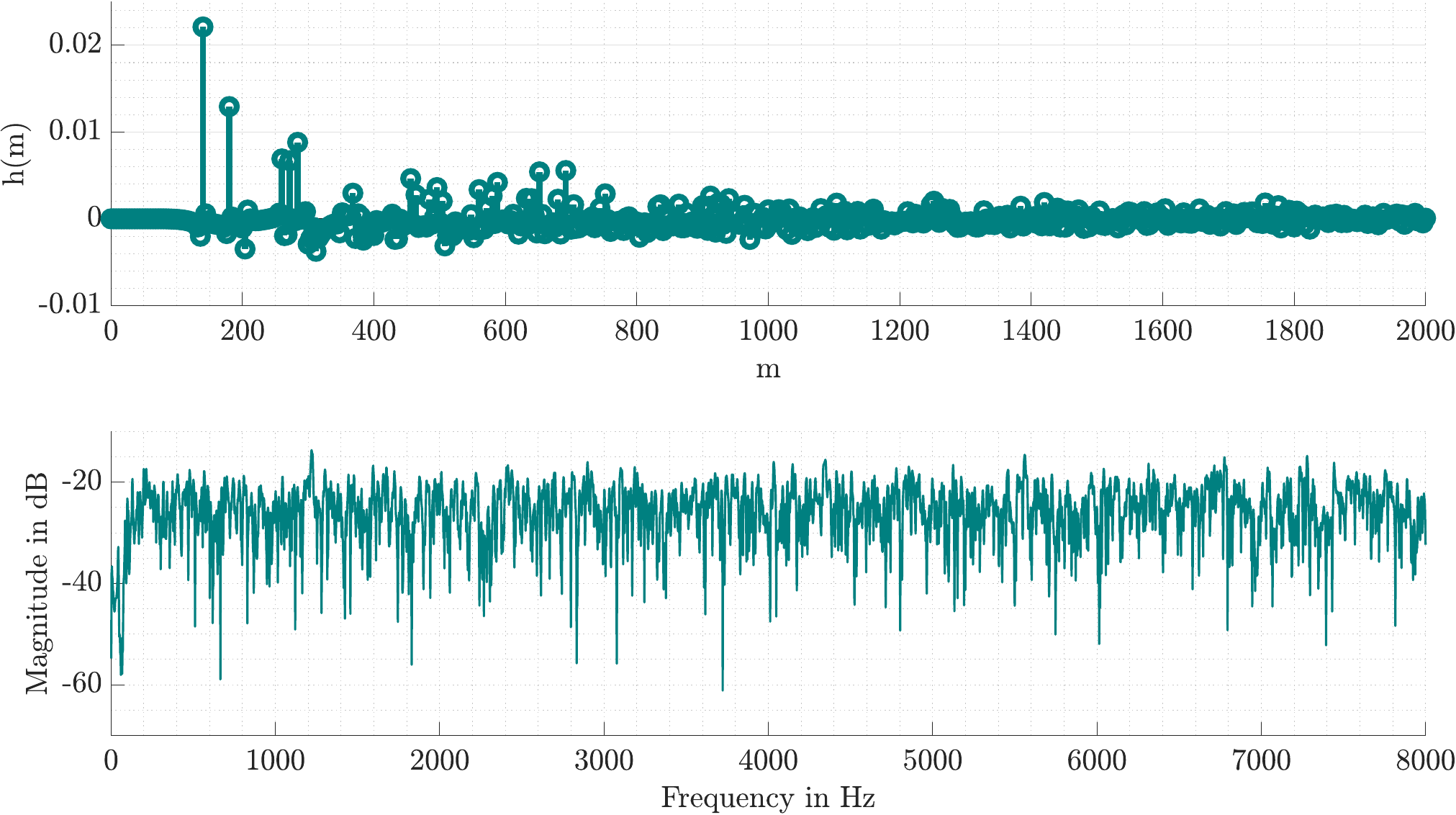}
    \caption{Simulated RIR for $\mathbf{b}= [8\text{\,m}, 7\text{\,m}, 2.8\text{\,m}]$, $\mathbf{s}=[3.9\text{\,m}, 3.4\text{\,m}, 1.2\text{\,m}]$ and $\mathbf{r}=[1.4\text{\,m}, 1.8\text{\,m}, 1.2\text{\,m}]$, and $T_{60} = 0.4 $ in the time domain (top) and the frequency domain (bottom).}
    \label{fig:rir}
\end{figure}

In order to sample random RIRs, we interpret these ten parameters to be random variables. We draw each value from a uniform distribution between a minimum and a maximum allowed value. For the room size and for $T_{60}$, the minimum and the maximum values can be chosen arbitrarily and are thus selected first. After those parameters are drawn, the ranges for source and receiver positions are drawn to guarantee that the source and the receiver are located inside the room. 

To simplify the notation, we use the 10-dimensional parameter vector $\theta$ in the following to describe all of these parameters. The RIR~$h$ can be considered as a sample of the distribution $ H_{\theta} $. An example of a simulated RIR in the time and the frequency domain is shown in Figure~\ref{fig:rir}. 
 
\subsection{Robust Audio Adversarial Examples}
Unlike earlier approaches that feed adversarial examples directly into the ASR system~\cite{Schoenherr2019, carlini2018audio}, we explicitly include characteristics of the room, in the form of RIRs, in the optimization problem. This hardens the adversarial examples to remain functional in an over-the-air~attack.

For the attack, we therefore extend the optimization criterion given in~\eqref{eq:optimization} by
\begin{equation}
x' = \arg \max_{\tilde{x}} \quad \mathbb{E}_{h\sim H_{\theta}}[P(y'|\tilde{x}_h)].
\label{eq:optrir}
\end{equation}
This approach is derived from the \emph{Expectation Over Transformation}~(EOT) approach in the visual domain, where it is used to consider different two- and three-dimensional transformations, which leads to robust real-world adversarial examples~\cite{athalye2017synthesizing}.
In our case, instead of visual transformations, we use the convolution with RIRs, drawn from $H_{\theta}$, to maximize the expectation over varying RIRs, as shown in Equation~\eqref{eq:optrir}.

\begin{figure}
    \centering
  \includegraphics[width=1.0\columnwidth]{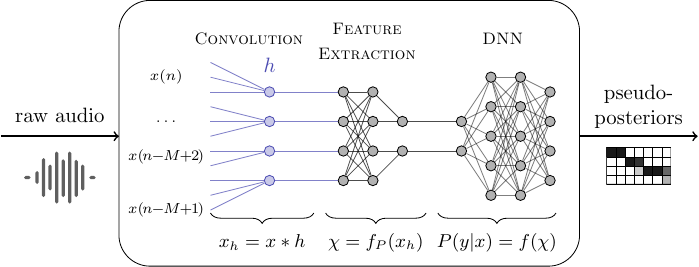}
    \caption{To simulate any RIR and to update the time domain audio signal directly, the RIR is integrated as an additional layer into the DNN.}
    \label{fig:dnn_detailed}
\end{figure}

For the implementation, we use a DNN that already has been augmented to include the feature extraction and prepend an additional layer to the DNN. This layer simulates the convolution from the input audio file with the RIR~$h$ to model the transmission through the room. Integrating this convolution as an additional layer allows us to apply gradient descent directly to the raw audio signal. 
A schematic overview of the proposed DNN is given in Figure~\ref{fig:dnn_detailed}. The first part (``Convolution'') describes the convolution with the RIR~$h$. Note that the RIR simulation layer is only used for the calculation of adversarial examples and removed during testing, as the actual  RIR will then act during the transmission over the air. The center and right part (``Feature extraction'' and ``DNN'') show the feature extraction and the acoustic model DNN, which is used to obtain the pseudo-posteriors for the decoding stage. 

The inclusion of the convolution as a layer in the DNN requires it to be differentiable. Using \eqref{eq:defconv}, the derivative can be written as
\begin{equation}
\dfrac{\partial x_h(n)}{\partial x(m)} = h(n-m) \quad \forall n,m.
\end{equation}
This can be integrated for the calculation of the gradient $\nabla x$ as
\begin{equation}
\nabla x = \dfrac{\partial L(y,y')}{\partial f(\chi)} \cdot \dfrac{\partial f(\chi)}{\partial f_p(x_h)} \cdot \dfrac{\partial f_p(x_h)}{\partial x_h} \cdot \dfrac{\partial x_h}{\partial x},
\end{equation}
where the function $f_p(\cdot)$ describes the feature extraction. This is an extension of the approach in~\cite{Schoenherr2019}, where
\begin{equation}
\nabla x = \dfrac{\partial L(y,y')}{\partial f(\chi)} \cdot \dfrac{\partial f(\chi)}{\partial f_p(x)} \cdot \dfrac{\partial f_p(x)}{\partial x}
\end{equation}
is defined for the calculation of adversarial examples via gradient descent using the objective function $L(\cdot)$. 

\subsection{Over-the-Air Adversarial Examples}
To assess the robustness of the hardened over-the-air adversarial attack, the adversarial examples~$x'$ have to be played back via a loudspeaker, and the recorded audio signals are used to determine the accuracy. 
%
% Algorithm
For the calculation, we implemented the optimization as defined in~\eqref{eq:optrir}, by sampling a new RIR~$h$ after every set of $Q$ gradient descent iterations. This simulates different rooms and recording conditions. Therefore, the generated adversarial example depends on the distribution $H_{\theta}$, from which the RIR~$h$ is drawn. After each gradient descent step, the audio signal~$x'$ is updated via the calculated gradient~$\nabla x$ at the learning rate~$\alpha$.

The total maximum number of iterations is limited to at most~$G$ iterations. However, if a successful robust adversarial example is created before the maximum number of iterations is reached, the algorithm does not need to continue. To efficiently calculate adversarial examples, we use an RIR~$h_{\text{test}}$ to simulate the over-the-air scenario during the calculation to verify whether the example has already achieved over-the-air robustness. This RIR is only used for verification and can, for example, be drawn out of $H_{\theta}$ once at the beginning of the algorithm.

The entire approach is summarized with Algorithm~\ref{alg:rir}. As can be seen, the psychoacoustic hearing thresholds~$\Phi$ are optionally used during the gradient descent to limit modifications of the signal to those time-frequency ranges, where they are (mostly) imperceptible. Here, $\text{DNN}_0$ describes the augmented DNN (``Feature extraction'' and ``DNN'') in Figure~\ref{fig:dnn} without the RIR simulation since, for the algorithm, this is replaced by the simulated RIR~$h_{\text{test}}$.

\begin{algorithm}[t]
    \caption{Calculation of robust adversarial examples.}
    \begin{algorithmic}[1]
    \State \textbf{input:} original audio $x$, target transcription $y'$, hearing thresholds~$\Phi$, distribution~$H_{\theta}$
    \State \textbf{result:} robust adversarial example $x'$ 
    \State \textbf{initialize:} $g \leftarrow 0$, $x' \leftarrow x$
    \While{$g < G \text{\textbf{ and }} y \neq y' $}
    \State $g \leftarrow g + 1$
    \State draw random sample $h \sim H_{\theta}$
    \State update first layer of DNN with $h$
      \For{$1$ \textbf{to} $Q$}
            \State // gradient descent, optionally constrained by
            \State //  hearing thresholds~$\Phi$

        \State $\nabla x \leftarrow  \frac{\partial L(y,y')}{\partial x} $
        \State $x' \leftarrow x' + \alpha \cdot \nabla x$
      \EndFor
    \State $x'_h \leftarrow x' * h_{\text{test}}$
    \State $y \leftarrow$ decode$(x'_h)$\ with $\text{DNN}_0$
    \EndWhile
    \end{algorithmic}
    \label{alg:rir}
\end{algorithm}

\section{Experimental Evaluation}

\begin{figure*}
    \centering
  	\includegraphics[width=2.0\columnwidth]{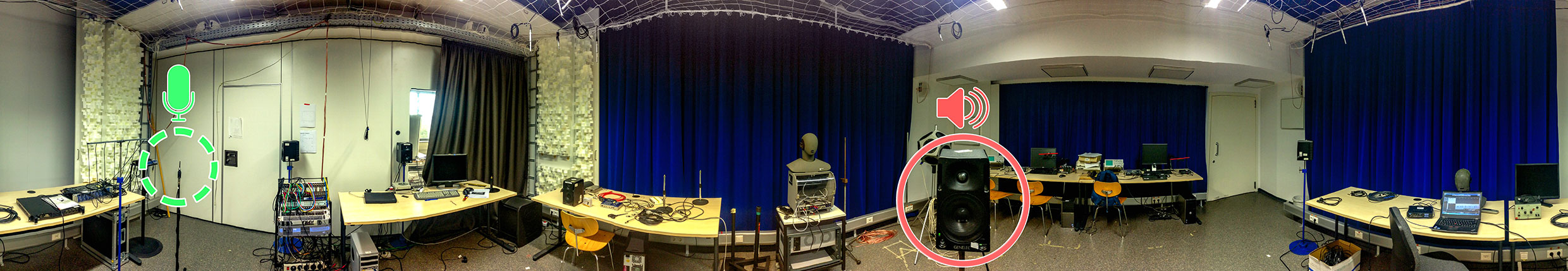}
    \caption{$360$ degree panorama shot of the lab setup used for the over-the-air recordings. The green dashed circle shows the microphone position and the red solid circle shows the position of the loudspeaker.}
    \label{fig:mmi}
\end{figure*}

In the following, we evaluate the performance of the proposed algorithm for adversarial examples played over-the-air and compare the performance for varying reverberation times, distances, and adversarial examples restricted by psychoacoustic hearing thresholds. 
Additionally, we compare the generic approach with an adapted version of the attack where an attacker has prior knowledge of the target room.
Finally, we measure the changes of generic adversarial examples replayed in different rooms and, even if there is no direct line-of-sight between the microphone and the speaker. 

For a practical demonstration of the attack, exemplary adversarial examples are available online at \camera{\url{http://imperio.adversarial-attacks.net}}.

\subsection{Metrics}
We use the following standard measures to assess the quality of the computed adversarial examples.

% WER
\subsubsection{Word Error Rate}
To measure performance, we use the word error rate (WER) with respect to the target transcription. For its computation, the standard metric for this purpose, the Levenshtein distance~\cite{navarro2001guided} $\mathcal{L}$, is used, summing up the number of deleted \(D\), inserted \(I\), and substituted \(S\) words for the best possible alignment between target text and recognition output. The Levenshtein distance is finally divided by the total number of words~\(N\) to obtain
\begin{equation}
WER  = 100 \cdot \frac{\mathcal{L}}{N} = 100 \cdot \frac{D + I + S}{N}.
\end{equation}
For a real attack, an adversarial example can only be considered successful if a WER of \(0\)\,\% is achieved (i.e., the hypothesis of the system matches with the attacker chosen target transcription).

% SNR
\subsubsection{Segmental Signal-to-Noise Ratio}
The segmental signal-to-noise ratio (SNRseg) measures the amount of noise~\(\sigma\) added from an attacker to the original signal~\(x\) and is computed as
\begin{equation}
\text{SNRseg(dB)} = \dfrac{10}{K} \sum_{k=0}^{K-1} \log _{10}  \frac{\sum_{t=Tk}^{Tk+T-1} x^2(t)}{\sum_{t=Tk}^{Tk+T-1} \sigma^2 (t)},
\end{equation}
where $T$ is the segment length and $K$ the number of segments. Thus, the higher the SNRseg, the \emph{less} noise was added. 

In contrast to the signal-to-noise ratio (SNR), the SNRseg~\cite{voran-1995-perception} is computed frame-wise and gives a better assessment of an audio signal if the original and the added noise are aligned~\cite{yang-1999-enhancedmb} as it is the case in our experiments.

\subsection{Calculation Time}

All experiments were performed on a machine with an Intel Xeon Gold 6130 CPU and 128~GB of DDR4 memory.

\begin{figure}[t]
    \centering
  	\includegraphics[width=1.\columnwidth]{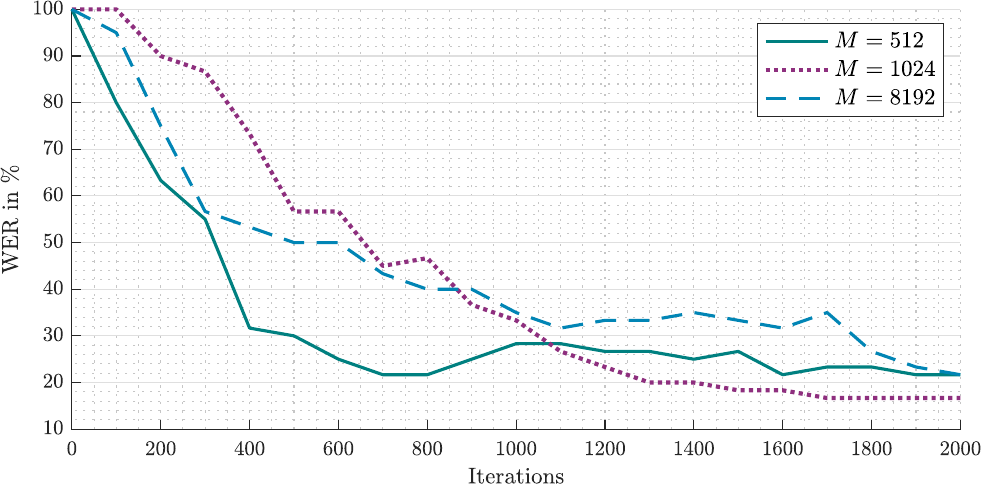}
    \caption{ WERs for simulated over-the-air attacks as a function of the maximum number of iterations $G$. }
    \label{fig:wer}
\end{figure}

For our experiments, we limit the maximum number of iterations to $2000$ since in every iteration more distortions are added to the audio file, which decreases the audio quality of the adversarial examples. Also, this number is sufficient for the attack to converge, as can be seen in Figure~\ref{fig:wer}, where the WER is plotted as a function of the maximum numbers of iterations~$G$. 

Computing an adversarial example for $10$ seconds of audio with the maximum number of $G = 2000$ iterations and $M = 512$ takes about $80$ minutes. Note that the computation for a single audio file is limited by the single-core performance of the machine, and the attack is fully parallelizable for multiple audio~files.

\subsection{Over-the-Air Attacks}
We evaluate the attack for the lab setup as shown in Figure~\ref{fig:mmi}. The approximate dimensions of this room are $\mathbf{b}_{\text{real}} \approx [8\text{\,m}, 7\text{\,m}, 2.8\text{\,m}]$ and the positions of the loudspeaker and the microphone are  $\mathbf{s}_{\text{real}}=[3.9\text{\,m}, 3.4\text{\,m}, 1.2\text{\,m}]$ and $\mathbf{r}_{\text{real}}=[1.4\text{\,m}, 1.8\text{\,m}, 1.2\text{\,m}]$, respectively. 

We compute all adversarial examples with Algorithm~\ref{alg:rir}. Based on preliminary experiments, we set $G = 2000$ and $Q = 10$. For the distributions~$H_{\theta}$, we used two different versions, shown in Table~\ref{tab:room}. $H_{\theta_{\text{gen}}}$ describes a generic room, while $H_{\theta_{\text{adp}}}$ is used as an approximation to reassemble the real room from Figure~\ref{fig:mmi}. If not specified otherwise, $h_{\text{test}}$, which is used for testing during the attack, is drawn once at the beginning of the algorithm from the same distribution $H_{\theta}$.

The WER is measured for the recorded adversarial examples after playing them via loudspeaker. The SNRseg is calculated after applying a measured RIR $h_{\text{real}}$ to both the original signal and the adversarial example. We chose this approach since it corresponds to the actual signal perceived by human listeners if the adversarial examples are played over the air. 

For all cases, we calculated $20$ adversarial examples. In some cases, the audio samples clipped too much (exceeded the maximum defined value of the audio, after the addition of the adversarial distortion). As it would not be possible to replay those examples, we removed them from the evaluation of the real over-the-air attack. Each of the remaining adversarial examples were played five times, and we reported the number of adversarial examples that could be transcribed with 0\,\% WER.

\subsubsection{Generic Over-the-Air Attack}
First, we evaluate the attack under the assumption that the attacker has no prior knowledge about the attack setup. Specifically, we use $H_{\theta_{\text{gen}}}$ and calculate adversarial examples for different reverberation times~$T_{60}$ and varying length $M$ of RIRs. $M$ describes how many past sampling values are considered, and the larger the reverberation time, the more important are the past sampling values.
We assume that, especially in setups with high reverberation times~$T_{60}$, a larger $M$ will result in more robust adversarial examples, as it is a better match to the real-world~conditions.

For the experiments, we used the variable-acoustics lab in Figure~\ref{fig:mmi} to adjust the reverberation time and tested three versions of the RIR length $M = 512$, $M = 1024$, and $M = 8192$ for speech data. The results in Table~\ref{tab:T60} confirm the above assumption: for $M = 8192$, we can obtain the best WERs, especially for the longer reverberation times. Note that even if the WER seems to be high, for an attacker, it is sufficient to play \emph{one} successful adversarial example with $0$\,\% WER, which is also in line with the definition in Section~\ref{sec:attack} and, in fact possible. The SNRseg decreases with increasing values for $M$, which indicates that more noise needs to be added to these adversarial examples.
However, the calculation time of the adversarial examples increases by the factor of four from $M = 512$ to $M = 8192$.

\begin{table}
\caption{Range of room dimensions for sampling the different distributions. $H_{\theta_{\text{gen}}}$ describes a generic room, which is used for the generic version of the attack, where we assume the attacker to have no prior knowledge. In case of $H_{\theta_{\text{adp}}}$, the distributions are adapted to the lab setup in Figure~\ref{fig:mmi}.}
\smallskip
\centering
\resizebox{1\columnwidth}{!}{
\begin{tabular}{l|cc|cc|cc|cc} 
\toprule
& \multicolumn{2}{c|}{$b_x$} & \multicolumn{2}{c|}{$b_y$} & \multicolumn{2}{c|}{$b_z$} & \multicolumn{2}{c}{$T_{60}$} \\
& min & max &  min & max & min & max & min & max \\
\midrule
$H_{\theta_{\text{gen}}}$  & 2.0\,m & 15.0\,m & 2.0\,m & 15.0\,m  & 2.0\,m & 5.0\,m  &  0.0\,s  &  1.0\,s \\
$H_{\theta_{\text{adp}}}$  & 6.0\,m & 10.0\,m & 5.0\,m & 9.0\,m  & 3.0\,m & 5.0\,m  & 0.2\,s & 0.6\,s \\ 
\bottomrule
\end{tabular}
}
\label{tab:room}
\end{table}

\begin{table}

\caption{WER, number of successful adversarial examples, and SNRseg for generic over-the-air attacks using $H_{\theta_{\text{gen}}}$ with speech data for different~$M$ and varying $T_{60}$.}
\centering
\smallskip
\normalsize
\resizebox{0.9\columnwidth}{!}{
\begin{tabular}{l|cc|cc|cc} 
\toprule
& \multicolumn{2}{c|}{$M = 512$} & \multicolumn{2}{c|}{$M = 1024$} & \multicolumn{2}{c}{$M = 8192$} \\ 
  & WER & AEs & WER & AEs & WER & AEs \\ 
\midrule 
$T_{60} = 0.42$\,s & 42.2\,\% & 5/20 & 34.9\,\% & 5/20 & 33.3\,\% & 2/20 \\
$T_{60} = 0.51$\,s & 68.9\,\% & 1/20 & 56.4\,\% & 2/20 & 42.0\,\% & 2/20 \\
$T_{60} = 0.65$\,s & ~91.6\,\% & 0/20 & 88.0\,\% & 0/20 & 68.7\,\% & 2/20\\
\midrule 
SNRseg & \multicolumn{2}{c|}{7.6$\pm$6.7\,dB} & \multicolumn{2}{c|}{7.7$\pm$6.7\,dB} & \multicolumn{2}{c}{3.2$\pm$6.1\,dB} \\
\bottomrule
\end{tabular}
}
\label{tab:T60}
\end{table}

\subsubsection{Hearing Thresholds}

To measure the impact of hearing thresholds, we conducted the same experiments as for Table~\ref{tab:T60} with $T_{60} = 0.42$ and hearing thresholds. The results are shown in Table~\ref{tab:thresholds}. Compared to the version without hearing thresholds, the WER and the total number of successful adversarial examples decrease. Nevertheless, it was possible to find successful adversarial examples for $M = 8192$. At the same time, the SNRseg has improved values. Additionally, the SNRseg measures any added noise, not only the perceptible noise components. Therefore, the perceptible noise is even lower than the SNRseg would suggest for the versions where hearing thresholds are used.

\begin{table}

\caption{WER, number of successful adversarial examples, and SNRseg for generic over-the-air attacks using $H_{\theta_{\text{gen}}}$ and hearing thresholds with speech data for different~$M$.}
\centering
\smallskip
\normalsize
\resizebox{.8\columnwidth}{!}{
\begin{tabular}{l|cc|cc|cc} 
\toprule
& \multicolumn{2}{c|}{$M = 512$} & \multicolumn{2}{c|}{$M = 1024$} & \multicolumn{2}{c}{$M = 8192$} \\ 
  & WER & AEs & WER & AEs & WER & AEs \\ 
\midrule 
$T_{60} = 0.42$\,s & 70.0\,\% & 0/20 &  62.7\,\% & 0/20 & 69.6\,\% & 2/20 \\
\midrule 
SNRseg & \multicolumn{2}{c|}{11.5$\pm$5.2\,dB} & \multicolumn{2}{c|}{10.4$\pm$6.9\,dB} & \multicolumn{2}{c}{5.5$\pm$4.8\,dB} \\
\bottomrule
\end{tabular}
}
\label{tab:thresholds}
\end{table}

\subsubsection{Distance between Speaker and Microphone}

\begin{figure}
    \centering
  	\includegraphics[width=1.0\columnwidth]{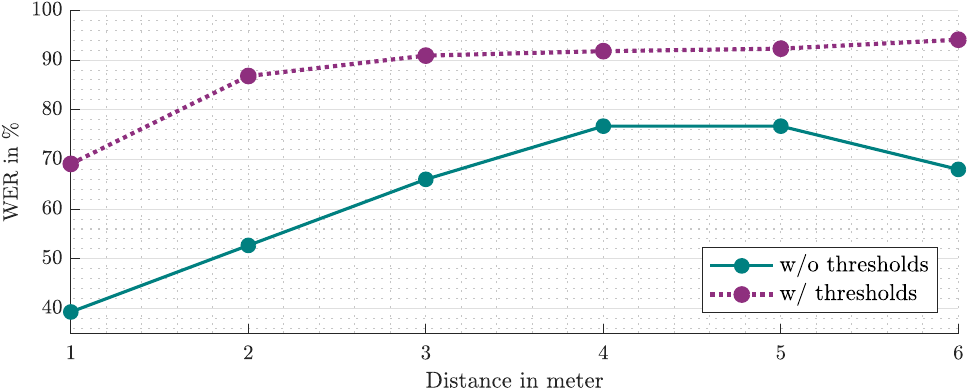}
    \caption{WERs for over-the-air attacks plotted over the distance between microphone and speaker for $M = 8192$ with and without hearing thresholds.}
    \label{fig:distance}
\end{figure}

In Figure~\ref{fig:distance}, we measured the effect of an increasing distance between the microphone and  loudspeaker. We used the shortest reverberation time $T_{60} = 0.42$\,s and varied the distance from $1$\,m to $6$\,m for $M = 8192$ with and without hearing thresholds.

In general, we find that the WER increases with increasing distance. Nevertheless, starting from a distance of approximately $2$\,m, the WER does not increase as rapidly as for smaller distances if we use hearing thresholds. In cases where no hearing thresholds are used, the WER even decreases for larger distances.

\subsubsection{Varying Audio Content}

In Table~\ref{tab:audio-content}, we evaluated the effect of varying audio contents of the original audio samples. For this, we used speech, music, and bird chirping data. Using speech audio samples for the attack results in the best WERs. 

The average SNRseg indicates that most distortions have to be added to the original audio samples for bird chirping while we achieve better results for music and speech data.

\subsubsection{Adaptive Attack}
In the following, we compare the generic attack, where the attacker has no prior knowledge about the attack setup, with an adapted version of the attack.
Note that the generic attack is the more powerful attack compared to the adapted version, as it requires no access nor any information about the room where the attack is conducted. 

For the evaluation, we used $H_{\theta_{\text{adp}}}$ and $H_{\theta_{\text{gen}}}$ in Table~\ref{tab:room}, combined with a measured RIR~$h_\text{real}$ and a simulated RIR~$h_\text{gen}$ for $h_\text{test}$ in Algorithm~\ref{alg:rir}. $h_\text{gen}$ was drawn once at the beginning of the algorithm from the same distribution, described via $H_{\theta_{\text{gen}}}$. $h_\text{real}$ is a measured RIR, obtained from the recording setup that is actually used during the attack.
Consequently, the version with $H_{\theta_{\text{gen}}}$ and $h_\text{gen}$ does not use \textit{any} prior knowledge of the room or the recording setup, while the version with $H_{\theta_{\text{adp}}}$ and $h_\text{real}$ is tailored to the room.
Surprisingly, the generic version clearly outperforms the adapted versions ($H_{\theta_{\text{adp}}}$, $h_\text{real}$) in Table~\ref{tab:generic-vs-adaptive}, and we were able to find fully successful adversarial examples for those cases, \ie adversarial examples with a WER of~$0$\,\%.

As a consequence, for an attacker, it is not only unnecessary to acquire prior knowledge about the room characteristics, but the likelihood of success is even higher if a generic attack is chosen.

\begin{table}
\caption{WER, number of successful adversarial examples, and SNRseg for different audio content for $M = 512$.}
\smallskip
\centering
\normalsize
\resizebox{.8\columnwidth}{!}{
\begin{tabular}{l|c|c|c}
\toprule
                & Music             & Speech            & Birds             \\
\midrule
Sucessful AEs   & 1/20              &  5/20             & 0/20              \\
WER             & 61.1\,\%         & 42.2\,\%         & 71.7\,\%         \\
\midrule
SNRseg          & 10.7$\pm$2.7\,dB  & ~7.6$\pm$6.7\,dB  & ~1.2$\pm$3.0\,dB  \\
\bottomrule
\end{tabular}
}
\label{tab:audio-content}
\end{table}

\begin{table}
\caption{WER, number of successful adversarial examples, and SNRseg for different audio content for $M = 512$. Comparing generic over-the-air attacks with adapted over-the-air attacks.}
\smallskip
\centering
\normalsize
\resizebox{.9\columnwidth}{!}{
\begin{tabular}{l|cc|cc|cc}
\toprule
                                & \multicolumn{2}{c|}{Music} & \multicolumn{2}{c|}{Speech} & \multicolumn{2}{c}{Birds} \\
                                  		& WER 		& AEs              & WER & AEs & WER & AEs	              \\
\midrule
$H_{\theta_\text{gen}}$, $h_\text{gen}$  & ~61.1\,\% & 1/20 & ~42.2\,\% & 5/20 & ~71.7\,\% & 0/20\\
$H_{\theta_\text{adp}}$, $h_\text{adp}$ & ~63.2\,\% & 2/20 & ~65.0\,\% & 2/20 & ~84.5\,\% & 2/20 \\
\midrule
$\Delta$ in WER & + ~2.1\,\% & &   + 22.8\,\% & &  + 12.8\,\% & \\
\bottomrule
\end{tabular}
}
\label{tab:generic-vs-adaptive}
\end{table}

\subsubsection{Varying Room Conditions}

\begin{table}
\caption{WER and number of successful adversarial examples for generic over-the-air attacks with and without direct line-of-sight in varying rooms based on speech data for~$M = 8192$.}
\smallskip
\centering
\normalsize
\resizebox{1.0\columnwidth}{!}{
\begin{tabular}{l|cc|cc|cc} 
\toprule
        & \multicolumn{2}{c|}{Lecture} 
                                & \multicolumn{2}{c|}{Meeting} 
                                                        & \multicolumn{2}{c}{\multirow{2}{*}{Office}}   \\ 
        & \multicolumn{2}{c|}{Room}    
                                & \multicolumn{2}{c|}{Room}    
                                                        &                                               \\
        & WER       & AEs       & WER       & AEs       & WER       & AEs                               \\
\midrule
w/\hspace{5pt} line-of-sight 
        & 40.0\,\%  & 2/20      & 55.3\,\%  & 1/20      & 74.0\,\%  & 1/20                              \\    
w/o line-of-sight 
        & 71.3\,\%  & 0/20      & 62.0\,\%  & 0/20      & 82.7\,\%  & 1/20                              \\  
\midrule
$\Delta$ in WER & + ~31.3\,\% & &   + ~6.7\,\% & &  + ~8,7\,\% & \\
\bottomrule
\end{tabular}
}
\label{tab:rooms}
\end{table}

To evaluate the adversarial examples in varying rooms, we chose three rooms of differing sizes: a lecture room with approximately $77$\,m$^2$, a meeting room with approximately $38$\,m$^2$, and an office with approximately $31$\,m$^2$. Layout plans of the rooms are shown in Appendix~\ref{app:rooms}, including positions of the speaker and microphone for all recording setups and the measured reverberation time.

\paragraph{Direct Line-of-Sight Attack}
The first attacks were conducted with a direct line-of-sight between the microphone and the speaker. The results are shown in Table~\ref{tab:rooms}. Even though the results vary depending on the room, the WERs remain approximately in the same range as the experiments with varying reverberation times in Table~\ref{tab:T60} would indicate. Surprisingly, the room with the highest reverberation time, the lecture room, actually gave the best results.

Overall, the results show that our generic adversarial examples remain robust for different kinds of rooms and setups and that it is sufficient to calculate one version of an adversarial example to cover a wide range of rooms (i.e., the attack is transferable).

\paragraph{No Line-of-Sight Attack}
For the rooms in Table~\ref{tab:rooms}, we also performed experiments where no line-of-sight between the microphone and the speaker exists by blocking the direct over-the-air connection with different kinds of furniture.
As a consequence, not the direct sound, but a reflected version of the audio is recorded.
An implication is that these attacks could be carried out without being visible to people in the vicinity of the ASR input microphone.
We tested different scenarios for our setup: In the lecture and the meeting room, the source and the receiver were separated by a table by simply placing the speaker under the table. In the office, the speaker was even placed \emph{outside} the room. For this recording setup, the door between the rooms was left open. For all other setups, the doors of the respective rooms were closed. A detailed description of the no line-of-sight setups is given in Appendix~\ref{app:rooms}. % The measured WERs for two different no-line-of-sight setups for each room are summarized in Table~\ref{tab:line-of-sight}.

In cases where no line-of-sight exists, the distortions that occur through the transmission can be considered more complex, and consequentially, a prediction of the recorded audio signal is hard. 
A blocked line-of-sight will most likely decrease the WER, but it is in general possible to find adversarial examples with $0$\,\% WER, even where the source was placed outside the room. This again shows that our generic version of the attack can successfully model a wide range of acoustic environments simultaneously, without any prior knowledge about the room~setup.

\section{Related Work}
In addition to the prior work that we have already discussed, we want to provide a broader and more detailed overview of related work in the following.

Generally speaking, adversarial attacks on ASR systems focus either on hiding a target transcription~\cite{carlini2016hidden, Abdullah-19-practical} or on obfuscating the original transcription~\cite{cisse2017houdini}. Almost all previous works on attacks against ASR systems did not focus on real-world attacks~\cite{carlini2016hidden,zhang2017dolphinattack} or were only successful for simulated over-the-air attacks~\cite{qin-19-robust}. 

Carlini~\etal~\cite{carlini2016hidden} have shown that targeted attacks against HMM-only ASR systems are possible. They use an inverse feature extraction to create adversarial audio samples. However, the resulting audio samples are not intelligible by humans in most cases and may be considered as noise, but may make thoughtful listeners suspicious once they are alerted to its hidden voice command.
An approach to overcome this limitation was proposed by Zhang~\etal~\cite{zhang2017dolphinattack}. They have shown that an adversary can hide a transcription by utilizing non-linearities of microphones to modulate the baseband audio signals with ultrasound above 20\,kHz, which they inject into the environment. The main downside of this attack, hence, is that the attacker needs to place an ultrasound transmitter in the vicinity of the voice-controlled system under attack and that the attacker needs to retrieve information from the audio signal, recorded with the specific microphone, which is costly in practice and tailors the attack to one specific setup.  Song and Mittael~\cite{song2017inaudible} and Roy \etal~\cite{roy2017backdoor} introduced similar ultrasound-based attacks that are not adversarial examples, but rather interact with the ASR system in a frequency range inaudible to humans. Nevertheless, for the attack hours of audio recordings are required to adjust the attack to the setup~\cite{song2017inaudible} or specially designed speakers are necessary~\cite{roy2017backdoor}.

Carlini and Wagner~\cite{carlini2018audio} published a general targeted attack on ASR systems using CTC-loss. The attacker creates the optimal attack via gradient-descent-based minimization~\cite{carlini2017towards} (similar to previous adversarial attacks on image classification), but the adversarial examples are fed directly into the recognizer. 
\emph{CommanderSong}~\cite{yuan-18-commandersong} is evaluated against \emph{Kaldi} and uses backpropagation to find an adversarial example. However, the very limited and non-systematic over-the-air attack highly depends on the speakers and recording devices, as the attack parameters have to be adjusted, especially for these components.
Yakura and Sakuma~\cite{yakura-2019-robust} published a technical report, which describes an algorithm to create over-the-air robust adversarial examples, but with the limitation that it is necessary to have physical access to the room where the attack takes place. Also, they did not evaluate their room-dependent results for varying room conditions and were unable to create generic adversarial examples systematically. Concurrently, Szuley and Kolter~\cite{szurley-2019-perceptual} also published a work on room-dependent robust adversarial examples, which worked under constraints given by a psychoacoustic model. However, their adversarial examples only work in an anechoic chamber, a room specifically designed to eliminate the effect of an RIR. The attack can, therefore, not be compared with a real-world scenario, as the anechoic chamber effectively reproduces the effect of directly feeding the attack into the ASR system, which is never given in real room environments.
Li~\etal~\cite{li-2019-adversarial-music} published a work to obfuscate Amazon's Alexa \emph{wake word} via specifically crafted music. However, their approach was not successful at creating \emph{targeted} adversarial examples that work over the air.

Alzantot~\etal~\cite{alzantot-2018-did} proposed a black-box attack, which does not require knowledge about the model. For this, the authors use a genetic algorithm to create their adversarial examples for a keyword spotting system, which differs from general speech recognition due to a much simpler architecture and far fewer possible recognition outcomes.
For \emph{DeepSpeech}~\cite{hannun-14-deepspeech} and \emph{Kaldi}, Khare~\etal~\cite{khare-19-blackbox} proposed a black-box attack based on evolutionary optimization, and also Taori~\etal~\cite{taori-2018-targeted} present a similar approach in their paper.

Sch\"onherr~\etal~\cite{Schoenherr2019} published an approach where psychoacoustic modeling, borrowed from the MP3 compression algorithm, was used to re-shape the perturbations of the adversarial examples in such a way as to hide the changes to the original audio below the human hearing thresholds. However, the adversarial examples created in that work need to be fed into the recognizer directly.
Concurrently, Abdullah~\etal~\cite{Abdullah-19-practical} showed a black-box attack in which psychoacoustics is used to calculate adversarial examples empirically. Their approach focuses on over-the-air attacks, but in many cases, humans can perceive the hidden message once they are alerted to its content. Note that our adversarial examples are conceptually completely different, as we use a target audio file, where we embed the target transcription via backpropagation. The changes, therefore, sound like random noise (see examples available at \camera{\url{http://imperio.adversarial-attacks.net}}). With Abdullah~\etal's approach, an audio file with the spoken target text is taken and changed in a way to be unintelligible for unbiased human listeners, but not for humans aware of the target transcription.
This is equally the case for Chen~\etal's~\cite{chen-2020-devil} recently published black-box attack against several commercial devices, where humans can perceive the target text.
As an extension of Carlini's and Wagner's attack~\cite{carlini2018audio}, Qin~\etal~\cite{qin-19-robust} introduced the first implementation of RIR-independent adversarial examples. Unfortunately, their approach only worked in a simulated environment and not for real over-the-air attacks, but the authors also utilize psychoacoustics to limit the~perturbations.

In the visual domain, Evtimov~\etal~\cite{evtimov2017robust} showed one of the first real-world adversarial attacks. They created and printed stickers, which can be used to obfuscate traffic signs. For humans, the stickers are visible. However, they seem very inconspicuous and could possibly fool autonomous cars.
Athalye~\etal~\cite{athalye2017synthesizing} presented another real-world adversarial perturbation on a 3D-printed turtle, which is recognized as a rifle from almost every point of view. The algorithm to create this 3D object not only minimizes the distortion for one image but for \emph{all} possible projections of a 3D object into a 2D image, hence producing a robust adversarial example.

Recently and independently, Chen~\etal~\cite{chen-2020-metamorph} showed a first over-the-air attack. Their attack was evaluated against \emph{DeepSpeech}~\cite{hannun-14-deepspeech}. Note that we focus on generic adversarial examples that work over-the-air for different kinds of rooms. Additionally, we used Kaldi, a hybrid DNN-HMM ASR system that works in a fundamentally different manner than the end-to-end approach of \emph{DeepSpeech}, which is attacked by Chen~\etal

Our approach is the first targeted attack that provides room-independent, robust adversarial examples against a hybrid ASR system. We demonstrate how to generate adversarial examples that are mostly unaffected by the environment, as ascertained by verifying their success in a broad range of room characteristics. We utilize the same psychoacoustics-based approach proposed by Sch\"onherr \etal~\cite{Schoenherr2019} to limit the perturbations of the audio signal to remain under, or at least close to, the human thresholds of hearing, and we show that the examples remain robust to playback over the air. The perturbations that remain audible in the adversarial examples that we create, are non-structured noise, so that human listeners cannot perceive any content related to the targeted recognition output.
Hence, our attack can be successful in a broad range of possible rooms, without any physical access to the environment (e.\,g., by playback of inconspicuous media from the Internet), and for which the target recognition output is not at all perceptible by human listeners. It shows the possibility and risk of a new attack vector, as no specialized hardware is needed for the playback and by being insensitive to the rooms in which the attacked systems are being~operated.

\section{Discussion}

Our experiments show that the adversarial examples, which we calculated with the proposed algorithm, remain robust even for high reverberation times or large distances between speaker and microphone. Also, the same adversarial examples can be successfully played over the air, even for setups where no direct line-of-sight~exists.

\textbf{Attack Parameters.}
Our comparison between the generic and the adapted version of the attack shows that the more powerful generic attack does not only have a similar success rate but can even outperform an adapted version where the attacker has prior knowledge of the target room. Consequently, an attacker only needs to calculate one generic adversarial example to cover a wide range of possible recording setups simultaneously.

For an attack, one successful adversarial example, which remains robust after being replayed (with a WER of $0$\,\%), is already enough. Therefore, the best strategy for an actual attack would be to calculate a set of adversarial examples containing the malicious transcription and to choose the most robust ones. In general, the results indicate a trade-off between the WER and the noise level: if no hearing thresholds are used, the WER is significantly better in comparison to examples with hearing thresholds. Nevertheless, even if the WER is better in cases without hearing thresholds, we have shown that it is indeed possible to calculate over-the-air-robust adversarial examples with hearing thresholds. Those adversarial examples contain less perceptible noise and are, therefore, less likely to be detected by human listeners. 
 
\textbf{End-to-end ASR systems.}
End-to-end ASR systems differ wide\-ly from the hybrid ASR systems used in this paper. However, the proposed attack only requires the possibility for backpropagation from the output to the input of the recognition network, and can therefore be applied to end-to-end systems. A simulated version of a similar attack with RIRs has been shown by Qin~\etal~\cite{qin-19-robust}. An adaptation of this attack is, therefore, most likely transferable to end-to-end ASR systems in the real world.

\textbf{Black-Box Attack.}
In a black-box scenario, the attacker has no access to the ASR system. However, even for this more challenging attack, it has been shown that it is possible to calculate adversarial examples, with the caveat that humans can perceive the hidden transcription if they are made aware of it~\cite{chen-2020-devil}.
The proposed approach is not easy to apply to black-box adversarial examples like commercial ASR systems such as Amazon's Alexa. Nevertheless, it should be feasible to use a similar approach in combination with a parameter-stealing attack~\cite{ilyas2017query,papernot2016transferability,tramer2016stealing,papernot2017blackbox,wang-18-hyperparameters}. Once the attacker can rebuild their own system, which reassembles the black-box system, the proposed algorithm can be used with that system as well.

\textbf{Countermeasures.}
% what are possible countermeasures?
To effectively prevent adversarial attacks, an ASR system needs either some kind of detection mechanism or needs to be hardened against adversarial examples. The detection of adversarial examples for known attacks might be feasible. However, no guarantees can be given against novel attacks in the long term. For this, it is necessary to build the ASR system to be adversarial-example-robust, \eg by mimicking the human perception of speech similar to images encoded in JPEG format~\cite{bafna-2018-fourier}. One step in this direction can be to focus the ASR system on only those signal components that are perceptible to the human listener and thus carry semantic information.

Additionally, not only the input data can be utilized to detect adversarial examples, but the ASR system's DNN can also serve this purpose. To achieve this, the uncertainty of the DNN estimation can be utilized to predict the reliability of the DNN output~\cite{gal-16-dropout, louizos-16-structured, lakshminarayanan-17-simple, daubener-2020-detecting}.
Due to the difficulty to creating robust adversarial example defenses, Carlini~\etal proposed a guideline for the evaluation of adversarial robustness, which lists all important properties of a successful countermeasure against adversarial examples~\cite{carlini-19-evaluating}.

\section{Conclusion}

In this paper, we have demonstrated that ASR systems are vulnerable against 
adversarial examples played over the air and we have
introduced an algorithm for the calculation of robust adversarial examples.  By simulating varying room setups, we can create highly robust adversarial examples that remain successful over the air in many environments. 

To substantiate our claims, we performed over-the-air attacks against \emph{Kaldi}; a state-of-the-art hybrid recognition framework that is used in Amazon's Alexa and other commercial ASR systems. 
We presented the results of empirical attacks for different room configurations. Our algorithm can be used with and without psychoacoustic hearing thresholds, limiting the perturbations to being less perceptible by humans. Furthermore, we have shown that it is possible to create targeted robust adversarial examples for varying rooms even if no direct line-of-sight between the microphone and the speakers exists, and even if the test room characteristics are completely unknown during the creation of the example.

Future work should investigate possible countermeasures such as using only the perceptible parts of the audio signal for recognition or using internal statistical information of the hybrid recognizer for detecting attacks.

\section*{Acknowledgments}
\camera{Funded by the Deutsche Forschungsgemeinschaft (DFG, German Research Foundation) under Germany's Excellence Strategy - EXC 2092 \textsc{CaSa} - 390781972.}

\bibliographystyle{ACM-Reference-Format}
\bibliography{references}


\begin{thebibliography}{41}

%%% the Web address.

\ifx \showCODEN    \undefined \def \showCODEN     #1{\unskip}     \fi
\ifx \showDOI      \undefined \def \showDOI       #1{#1}\fi
\ifx \showISBNx    \undefined \def \showISBNx     #1{\unskip}     \fi
\ifx \showISBNxiii \undefined \def \showISBNxiii  #1{\unskip}     \fi
\ifx \showISSN     \undefined \def \showISSN      #1{\unskip}     \fi
\ifx \showLCCN     \undefined \def \showLCCN      #1{\unskip}     \fi
\ifx \shownote     \undefined \def \shownote      #1{#1}          \fi
\ifx \showarticletitle \undefined \def \showarticletitle #1{#1}   \fi
\ifx \showURL      \undefined \def \showURL       {\relax}        \fi
% invisible to TeX
\providecommand\bibfield[2]{#2}
\providecommand\bibinfo[2]{#2}
\providecommand\natexlab[1]{#1}
\providecommand\showeprint[2][]{arXiv:#2}

\bibitem[\protect\citeauthoryear{Abdullah, Garcia, Peeters, Traynor, Butler,
  and Wilson}{Abdullah et~al\mbox{.}}{2019}]%
        {Abdullah-19-practical}
\bibfield{author}{\bibinfo{person}{Hadi Abdullah}, \bibinfo{person}{Washington
  Garcia}, \bibinfo{person}{Christian Peeters}, \bibinfo{person}{Patrick
  Traynor}, \bibinfo{person}{Kevin R.~B. Butler}, {and} \bibinfo{person}{Joseph
  Wilson}.} \bibinfo{year}{2019}\natexlab{}.
\newblock \showarticletitle{Practical Hidden Voice Attacks against Speech and
  Speaker Recognition Systems}. In \bibinfo{booktitle}{\emph{Network and
  Distributed System Security Symposium (NDSS)}}.
\newblock


\bibitem[\protect\citeauthoryear{Allen and Berkley}{Allen and Berkley}{1979}]%
        {allen-1979-image}
\bibfield{author}{\bibinfo{person}{Jont~B. Allen} {and}
  \bibinfo{person}{David~A. Berkley}.} \bibinfo{year}{1979}\natexlab{}.
\newblock \showarticletitle{Image method for efficiently simulating small-room
  acoustics}.
\newblock \bibinfo{journal}{\emph{The Journal of the Acoustical Society of
  America}} \bibinfo{volume}{65}, \bibinfo{number}{4} (\bibinfo{year}{1979}),
  \bibinfo{pages}{943--950}.
\newblock


\bibitem[\protect\citeauthoryear{Alzantot, Balaji, and Srivastava}{Alzantot
  et~al\mbox{.}}{2018}]%
        {alzantot-2018-did}
\bibfield{author}{\bibinfo{person}{Moustafa Alzantot},
  \bibinfo{person}{Bharathan Balaji}, {and} \bibinfo{person}{Mani Srivastava}.}
  \bibinfo{year}{2018}\natexlab{}.
\newblock \showarticletitle{Did you hear that? {A}dversarial examples against
  automatic speech recognition}.
\newblock \bibinfo{journal}{\emph{arXiv preprint arXiv:1801.00554}}
  (\bibinfo{year}{2018}).
\newblock


\bibitem[\protect\citeauthoryear{Athalye, Engstrom, Ilyas, and Kwok}{Athalye
  et~al\mbox{.}}{2017}]%
        {athalye2017synthesizing}
\bibfield{author}{\bibinfo{person}{Anish Athalye}, \bibinfo{person}{Logan
  Engstrom}, \bibinfo{person}{Andrew Ilyas}, {and} \bibinfo{person}{Kevin
  Kwok}.} \bibinfo{year}{2017}\natexlab{}.
\newblock \showarticletitle{Synthesizing Robust Adversarial Examples}.
\newblock \bibinfo{journal}{\emph{CoRR}}  \bibinfo{volume}{abs/1707.07397}
  (\bibinfo{date}{July} \bibinfo{year}{2017}), \bibinfo{pages}{1--18}.
\newblock


\bibitem[\protect\citeauthoryear{Bafna, Murtagh, and Vyas}{Bafna
  et~al\mbox{.}}{2018}]%
        {bafna-2018-fourier}
\bibfield{author}{\bibinfo{person}{Mitali Bafna}, \bibinfo{person}{Jack
  Murtagh}, {and} \bibinfo{person}{Nikhil Vyas}.}
  \bibinfo{year}{2018}\natexlab{}.
\newblock \showarticletitle{Thwarting Adversarial Examples: An {${L}_1$}-Robust
  Sparse Fourier Transform}. In \bibinfo{booktitle}{\emph{Advances in Neural
  Information Processing Systems 31}}. \bibinfo{pages}{10075--10085}.
\newblock


\bibitem[\protect\citeauthoryear{Carlini, Athalye, Papernot, Brendel, Rauber,
  Tsipras, Goodfellow, and Madry}{Carlini et~al\mbox{.}}{2019}]%
        {carlini-19-evaluating}
\bibfield{author}{\bibinfo{person}{Nicholas Carlini}, \bibinfo{person}{Anish
  Athalye}, \bibinfo{person}{Nicolas Papernot}, \bibinfo{person}{Wieland
  Brendel}, \bibinfo{person}{Jonas Rauber}, \bibinfo{person}{Dimitris Tsipras},
  \bibinfo{person}{Ian Goodfellow}, {and} \bibinfo{person}{Aleksander Madry}.}
  \bibinfo{year}{2019}\natexlab{}.
\newblock \showarticletitle{On evaluating adversarial robustness}.
\newblock \bibinfo{journal}{\emph{arXiv preprint arXiv:1902.06705}}
  (\bibinfo{year}{2019}).
\newblock


\bibitem[\protect\citeauthoryear{Carlini, Mishra, Vaidya, Zhang, Sherr,
  Shields, Wagner, and Zhou}{Carlini et~al\mbox{.}}{2016}]%
        {carlini2016hidden}
\bibfield{author}{\bibinfo{person}{Nicholas Carlini}, \bibinfo{person}{Pratyush
  Mishra}, \bibinfo{person}{Tavish Vaidya}, \bibinfo{person}{Yuankai Zhang},
  \bibinfo{person}{Micah Sherr}, \bibinfo{person}{Clay Shields},
  \bibinfo{person}{David~A. Wagner}, {and} \bibinfo{person}{Wenchao Zhou}.}
  \bibinfo{year}{2016}\natexlab{}.
\newblock \showarticletitle{Hidden Voice Commands}. In
  \bibinfo{booktitle}{\emph{USENIX Security Symposium}}.
  \bibinfo{publisher}{USENIX}, \bibinfo{pages}{513--530}.
\newblock


\bibitem[\protect\citeauthoryear{Carlini and Wagner}{Carlini and
  Wagner}{2017}]%
        {carlini2017towards}
\bibfield{author}{\bibinfo{person}{Nicholas Carlini} {and}
  \bibinfo{person}{David Wagner}.} \bibinfo{year}{2017}\natexlab{}.
\newblock \showarticletitle{Towards Evaluating the Robustness of Neural
  Networks}. In \bibinfo{booktitle}{\emph{Symposium on Security and Privacy}}.
  \bibinfo{publisher}{IEEE}, \bibinfo{pages}{39--57}.
\newblock


\bibitem[\protect\citeauthoryear{Carlini and Wagner}{Carlini and
  Wagner}{2018}]%
        {carlini2018audio}
\bibfield{author}{\bibinfo{person}{Nicholas Carlini} {and}
  \bibinfo{person}{David Wagner}.} \bibinfo{year}{2018}\natexlab{}.
\newblock \showarticletitle{Audio adversarial examples: Targeted attacks on
  speech-to-text}.
\newblock  (\bibinfo{year}{2018}), \bibinfo{pages}{1--7}.
\newblock


\bibitem[\protect\citeauthoryear{Chen, Shangguan, Li, and Jamieson}{Chen
  et~al\mbox{.}}{2020a}]%
        {chen-2020-metamorph}
\bibfield{author}{\bibinfo{person}{Tao Chen}, \bibinfo{person}{Longfei
  Shangguan}, \bibinfo{person}{Zhenjiang Li}, {and} \bibinfo{person}{Kyle
  Jamieson}.} \bibinfo{year}{2020}\natexlab{a}.
\newblock \showarticletitle{Metamorph: Injecting Inaudible Commands into
  Over-the-air Voice Controlled Systems}.
\newblock  (\bibinfo{year}{2020}).
\newblock


\bibitem[\protect\citeauthoryear{Chen, Yuan, Zhang, Zhao, Zhang, Chen, and
  Wang}{Chen et~al\mbox{.}}{2020b}]%
        {chen-2020-devil}
\bibfield{author}{\bibinfo{person}{Yuxuan Chen}, \bibinfo{person}{Xuejing
  Yuan}, \bibinfo{person}{Jiangshan Zhang}, \bibinfo{person}{Yue Zhao},
  \bibinfo{person}{Shengzhi Zhang}, \bibinfo{person}{Kai Chen}, {and}
  \bibinfo{person}{XiaoFeng Wang}.} \bibinfo{year}{2020}\natexlab{b}.
\newblock \showarticletitle{Devil’s Whisper: A General Approach for Physical
  Adversarial Attacks against Commercial Black-box Speech Recognition Devices}.
  In \bibinfo{booktitle}{\emph{USENIX Security Symposium}}.
  \bibinfo{publisher}{USENIX}.
\newblock


\bibitem[\protect\citeauthoryear{Cisse, Adi, Neverova, and Keshet}{Cisse
  et~al\mbox{.}}{2017}]%
        {cisse2017houdini}
\bibfield{author}{\bibinfo{person}{Moustapha Cisse}, \bibinfo{person}{Yossi
  Adi}, \bibinfo{person}{Natalia Neverova}, {and} \bibinfo{person}{Joseph
  Keshet}.} \bibinfo{year}{2017}\natexlab{}.
\newblock \showarticletitle{{Houdini}: {F}ooling Deep Structured Prediction
  Models}.
\newblock \bibinfo{journal}{\emph{CoRR}}  \bibinfo{volume}{abs/1707.05373}
  (\bibinfo{date}{July} \bibinfo{year}{2017}), \bibinfo{pages}{1--12}.
\newblock


\bibitem[\protect\citeauthoryear{D{\"a}ubener, Sch{\"o}nherr, Fischer, and
  Kolossa}{D{\"a}ubener et~al\mbox{.}}{2020}]%
        {daubener-2020-detecting}
\bibfield{author}{\bibinfo{person}{Sina D{\"a}ubener}, \bibinfo{person}{Lea
  Sch{\"o}nherr}, \bibinfo{person}{Asja Fischer}, {and}
  \bibinfo{person}{Dorothea Kolossa}.} \bibinfo{year}{2020}\natexlab{}.
\newblock \showarticletitle{Detecting Adversarial Examples for Speech
  Recognition via Uncertainty Quantification}.
\newblock \bibinfo{journal}{\emph{arXiv preprint arXiv:2005.14611}}
  (\bibinfo{year}{2020}).
\newblock


\bibitem[\protect\citeauthoryear{Evtimov, Eykholt, Fernandes, Kohno, Li,
  Prakash, Rahmati, and Song}{Evtimov et~al\mbox{.}}{2017}]%
        {evtimov2017robust}
\bibfield{author}{\bibinfo{person}{Ivan Evtimov}, \bibinfo{person}{Kevin
  Eykholt}, \bibinfo{person}{Earlence Fernandes}, \bibinfo{person}{Tadayoshi
  Kohno}, \bibinfo{person}{Bo Li}, \bibinfo{person}{Atul Prakash},
  \bibinfo{person}{Amir Rahmati}, {and} \bibinfo{person}{Dawn Song}.}
  \bibinfo{year}{2017}\natexlab{}.
\newblock \showarticletitle{Robust Physical-World Attacks on Machine Learning
  Models}.
\newblock \bibinfo{journal}{\emph{CoRR}}  \bibinfo{volume}{abs/1707.08945}
  (\bibinfo{date}{July} \bibinfo{year}{2017}), \bibinfo{pages}{1--11}.
\newblock


\bibitem[\protect\citeauthoryear{Gal and Ghahramani}{Gal and
  Ghahramani}{2016}]%
        {gal-16-dropout}
\bibfield{author}{\bibinfo{person}{Yarin Gal} {and} \bibinfo{person}{Zoubin
  Ghahramani}.} \bibinfo{year}{2016}\natexlab{}.
\newblock \showarticletitle{Dropout as a bayesian approximation: Representing
  model uncertainty in deep learning}. In
  \bibinfo{booktitle}{\emph{International Conference on Machine Learning}}.
  \bibinfo{pages}{1050--1059}.
\newblock


\bibitem[\protect\citeauthoryear{Hannun, Case, Casper, Catanzaro, Diamos,
  Elsen, Prenger, Satheesh, Sengupta, Coates, et~al\mbox{.}}{Hannun
  et~al\mbox{.}}{2014}]%
        {hannun-14-deepspeech}
\bibfield{author}{\bibinfo{person}{Awni Hannun}, \bibinfo{person}{Carl Case},
  \bibinfo{person}{Jared Casper}, \bibinfo{person}{Bryan Catanzaro},
  \bibinfo{person}{Greg Diamos}, \bibinfo{person}{Erich Elsen},
  \bibinfo{person}{Ryan Prenger}, \bibinfo{person}{Sanjeev Satheesh},
  \bibinfo{person}{Shubho Sengupta}, \bibinfo{person}{Adam Coates},
  {et~al\mbox{.}}} \bibinfo{year}{2014}\natexlab{}.
\newblock \showarticletitle{Deep speech: Scaling up end-to-end speech
  recognition}.
\newblock \bibinfo{journal}{\emph{arXiv preprint arXiv:1412.5567}}
  (\bibinfo{year}{2014}).
\newblock


\bibitem[\protect\citeauthoryear{Ilyas, Engstrom, Athalye, and Lin}{Ilyas
  et~al\mbox{.}}{2018}]%
        {ilyas2017query}
\bibfield{author}{\bibinfo{person}{Andrew Ilyas}, \bibinfo{person}{Logan
  Engstrom}, \bibinfo{person}{Anish Athalye}, {and} \bibinfo{person}{Jessy
  Lin}.} \bibinfo{year}{2018}\natexlab{}.
\newblock \showarticletitle{{Black-box} Adversarial Attacks with Limited
  Queries and Information}.
\newblock \bibinfo{journal}{\emph{CoRR}}  \bibinfo{volume}{abs/1804.08598}
  (\bibinfo{date}{April} \bibinfo{year}{2018}), \bibinfo{pages}{1--10}.
\newblock


\bibitem[\protect\citeauthoryear{ISO}{ISO}{1993}]%
        {ISO11172}
\bibfield{author}{\bibinfo{person}{ISO}.} \bibinfo{year}{1993}\natexlab{}.
\newblock \bibinfo{booktitle}{\emph{Information {T}echnology -- {C}oding of
  Moving Pictures and Associated Audio for Digital Storage Media at Up to 1.5
  {Mbits/s} -- {Part3}: {A}udio}}.
\newblock \bibinfo{type}{ISO} 11172-3. \bibinfo{institution}{International
  Organization for Standardization}.
\newblock


\bibitem[\protect\citeauthoryear{Lakshminarayanan, Pritzel, and
  Blundell}{Lakshminarayanan et~al\mbox{.}}{2017}]%
        {lakshminarayanan-17-simple}
\bibfield{author}{\bibinfo{person}{Balaji Lakshminarayanan},
  \bibinfo{person}{Alexander Pritzel}, {and} \bibinfo{person}{Charles
  Blundell}.} \bibinfo{year}{2017}\natexlab{}.
\newblock \showarticletitle{Simple and scalable predictive uncertainty
  estimation using deep ensembles}. In \bibinfo{booktitle}{\emph{Advances in
  Neural Information Processing Systems}}. \bibinfo{pages}{6402--6413}.
\newblock


\bibitem[\protect\citeauthoryear{Li, Qu, Li, Szurley, Kolter, and Metze}{Li
  et~al\mbox{.}}{2019}]%
        {li-2019-adversarial-music}
\bibfield{author}{\bibinfo{person}{Juncheng Li}, \bibinfo{person}{Shuhui Qu},
  \bibinfo{person}{Xinjian Li}, \bibinfo{person}{Joseph Szurley},
  \bibinfo{person}{J~Zico Kolter}, {and} \bibinfo{person}{Florian Metze}.}
  \bibinfo{year}{2019}\natexlab{}.
\newblock \showarticletitle{Adversarial Music: Real World Audio Adversary
  Against Wake-word Detection System}. In \bibinfo{booktitle}{\emph{Advances in
  Neural Information Processing Systems (NeurIPS)}}.
  \bibinfo{pages}{11908--11918}.
\newblock


\bibitem[\protect\citeauthoryear{Louizos and Welling}{Louizos and
  Welling}{2016}]%
        {louizos-16-structured}
\bibfield{author}{\bibinfo{person}{Christos Louizos} {and} \bibinfo{person}{Max
  Welling}.} \bibinfo{year}{2016}\natexlab{}.
\newblock \showarticletitle{Structured and efficient variational deep learning
  with matrix gaussian posteriors}. In \bibinfo{booktitle}{\emph{International
  Conference on Machine Learning}}. \bibinfo{pages}{1708--1716}.
\newblock


\bibitem[\protect\citeauthoryear{L{\"u}scher, Beck, Irie, Kitza, Michel, Zeyer,
  Schl{\"u}ter, and Ney}{L{\"u}scher et~al\mbox{.}}{2019}]%
        {luscher-19-rwth}
\bibfield{author}{\bibinfo{person}{Christoph L{\"u}scher},
  \bibinfo{person}{Eugen Beck}, \bibinfo{person}{Kazuki Irie},
  \bibinfo{person}{Markus Kitza}, \bibinfo{person}{Wilfried Michel},
  \bibinfo{person}{Albert Zeyer}, \bibinfo{person}{Ralf Schl{\"u}ter}, {and}
  \bibinfo{person}{Hermann Ney}.} \bibinfo{year}{2019}\natexlab{}.
\newblock \showarticletitle{{RWTH ASR} systems for LibriSpeech: Hybrid vs
  Attention}.
\newblock \bibinfo{journal}{\emph{Proceedings of Interspeech}}
  (\bibinfo{year}{2019}), \bibinfo{pages}{231--235}.
\newblock


\bibitem[\protect\citeauthoryear{Navarro}{Navarro}{2001}]%
        {navarro2001guided}
\bibfield{author}{\bibinfo{person}{Gonzalo Navarro}.}
  \bibinfo{year}{2001}\natexlab{}.
\newblock \showarticletitle{A Guided Tour to Approximate String Matching}.
\newblock \bibinfo{journal}{\emph{Comput. Surveys}} \bibinfo{volume}{33},
  \bibinfo{number}{1} (\bibinfo{date}{March} \bibinfo{year}{2001}),
  \bibinfo{pages}{31--88}.
\newblock


\bibitem[\protect\citeauthoryear{Papernot, McDaniel, Goodfellow, Jha, Celik,
  and Swami}{Papernot et~al\mbox{.}}{2017}]%
        {papernot2017blackbox}
\bibfield{author}{\bibinfo{person}{Nicolas Papernot}, \bibinfo{person}{Patrick
  McDaniel}, \bibinfo{person}{Ian Goodfellow}, \bibinfo{person}{Somesh Jha},
  \bibinfo{person}{Z.~Berkay Celik}, {and} \bibinfo{person}{Ananthram Swami}.}
  \bibinfo{year}{2017}\natexlab{}.
\newblock \showarticletitle{Practical {Black-Box} Attacks Against Machine
  Learning}. In \bibinfo{booktitle}{\emph{Asia Conference on Computer and
  Communications Security (ASIA CCS)}}. \bibinfo{publisher}{ACM},
  \bibinfo{pages}{506--519}.
\newblock


\bibitem[\protect\citeauthoryear{Papernot, McDaniel, and Goodfellow}{Papernot
  et~al\mbox{.}}{2016}]%
        {papernot2016transferability}
\bibfield{author}{\bibinfo{person}{Nicolas Papernot},
  \bibinfo{person}{Patrick~D. McDaniel}, {and} \bibinfo{person}{Ian~J.
  Goodfellow}.} \bibinfo{year}{2016}\natexlab{}.
\newblock \showarticletitle{Transferability in Machine Learning: {F}rom
  Phenomena to {Black-Box} Attacks using Adversarial Samples}.
\newblock \bibinfo{journal}{\emph{CoRR}}  \bibinfo{volume}{abs/1605.07277}
  (\bibinfo{date}{May} \bibinfo{year}{2016}), \bibinfo{pages}{1--13}.
\newblock


\bibitem[\protect\citeauthoryear{Povey, Ghoshal, Boulianne, Burget, Glembek,
  Goel, Hannemann, Motlicek, Qian, Schwarz, Silovsky, Stemmer, and
  Vesely}{Povey et~al\mbox{.}}{2011}]%
        {Povey_ASRU2011_2011}
\bibfield{author}{\bibinfo{person}{Daniel Povey}, \bibinfo{person}{Arnab
  Ghoshal}, \bibinfo{person}{Gilles Boulianne}, \bibinfo{person}{Lukas Burget},
  \bibinfo{person}{Ondrej Glembek}, \bibinfo{person}{Nagendra Goel},
  \bibinfo{person}{Mirko Hannemann}, \bibinfo{person}{Petr Motlicek},
  \bibinfo{person}{Yanmin Qian}, \bibinfo{person}{Petr Schwarz},
  \bibinfo{person}{Jan Silovsky}, \bibinfo{person}{Georg Stemmer}, {and}
  \bibinfo{person}{Karel Vesely}.} \bibinfo{year}{2011}\natexlab{}.
\newblock \showarticletitle{The {Kaldi} Speech Recognition Toolkit}. In
  \bibinfo{booktitle}{\emph{Workshop on Automatic Speech Recognition and
  Understanding}}. \bibinfo{publisher}{IEEE}.
\newblock


\bibitem[\protect\citeauthoryear{Qin, Carlini, Goodfellow, Cottrell, and
  Raffel}{Qin et~al\mbox{.}}{2019}]%
        {qin-19-robust}
\bibfield{author}{\bibinfo{person}{Yao Qin}, \bibinfo{person}{Nicholas
  Carlini}, \bibinfo{person}{Ian Goodfellow}, \bibinfo{person}{Garrison
  Cottrell}, {and} \bibinfo{person}{Colin Raffel}.}
  \bibinfo{year}{2019}\natexlab{}.
\newblock \showarticletitle{Imperceptible, Robust, and Targeted Adversarial
  Examples for Automatic Speech Recognition}. In
  \bibinfo{booktitle}{\emph{arXiv preprint arXiv:1903.10346}}.
\newblock


\bibitem[\protect\citeauthoryear{Roy, Hassanieh, and Roy~Choudhury}{Roy
  et~al\mbox{.}}{2017}]%
        {roy2017backdoor}
\bibfield{author}{\bibinfo{person}{Nirupam Roy}, \bibinfo{person}{Haitham
  Hassanieh}, {and} \bibinfo{person}{Romit Roy~Choudhury}.}
  \bibinfo{year}{2017}\natexlab{}.
\newblock \showarticletitle{{BackDoor}: {M}aking Microphones Hear Inaudible
  Sounds}. In \bibinfo{booktitle}{\emph{Conference on Mobile Systems,
  Applications, and Services}}. \bibinfo{publisher}{ACM},
  \bibinfo{pages}{2--14}.
\newblock


\bibitem[\protect\citeauthoryear{Sch\"{o}nherr, Kohls, Zeiler, Holz, and
  Kolossa}{Sch\"{o}nherr et~al\mbox{.}}{2019}]%
        {Schoenherr2019}
\bibfield{author}{\bibinfo{person}{Lea Sch\"{o}nherr},
  \bibinfo{person}{Katharina Kohls}, \bibinfo{person}{Steffen Zeiler},
  \bibinfo{person}{Thorsten Holz}, {and} \bibinfo{person}{Dorothea Kolossa}.}
  \bibinfo{year}{2019}\natexlab{}.
\newblock \showarticletitle{Adversarial Attacks Against Automatic Speech
  Recognition Systems via Psychoacoustic Hiding}. In
  \bibinfo{booktitle}{\emph{Network and Distributed System Security Symposium
  (NDSS)}}.
\newblock


\bibitem[\protect\citeauthoryear{Shreya~Khare}{Shreya~Khare}{2019}]%
        {khare-19-blackbox}
\bibfield{author}{\bibinfo{person}{Senthil~Mani Shreya~Khare,
  Rahul~Aralikatte}.} \bibinfo{year}{2019}\natexlab{}.
\newblock \showarticletitle{Adversarial Black-Box Attacks on Automatic Speech
  Recognition Systems using Multi-Objective Evolutionary Optimization}.
\newblock \bibinfo{journal}{\emph{Proceedings of Interspeech}}
  (\bibinfo{year}{2019}).
\newblock


\bibitem[\protect\citeauthoryear{Song and Mittal}{Song and Mittal}{2017}]%
        {song2017inaudible}
\bibfield{author}{\bibinfo{person}{Liwei Song} {and} \bibinfo{person}{Prateek
  Mittal}.} \bibinfo{year}{2017}\natexlab{}.
\newblock \showarticletitle{Inaudible Voice Commands}.
\newblock \bibinfo{journal}{\emph{CoRR}}  \bibinfo{volume}{abs/1708.07238}
  (\bibinfo{date}{Aug.} \bibinfo{year}{2017}), \bibinfo{pages}{1--3}.
\newblock


\bibitem[\protect\citeauthoryear{Szurley and Kolter}{Szurley and
  Kolter}{2019}]%
        {szurley-2019-perceptual}
\bibfield{author}{\bibinfo{person}{Joseph Szurley} {and}
  \bibinfo{person}{J~Zico Kolter}.} \bibinfo{year}{2019}\natexlab{}.
\newblock \showarticletitle{Perceptual Based Adversarial Audio Attacks}.
\newblock \bibinfo{journal}{\emph{arXiv preprint arXiv:1906.06355}}
  (\bibinfo{year}{2019}).
\newblock


\bibitem[\protect\citeauthoryear{Taori, Kamsetty, Chu, and Vemuri}{Taori
  et~al\mbox{.}}{2018}]%
        {taori-2018-targeted}
\bibfield{author}{\bibinfo{person}{Rohan Taori}, \bibinfo{person}{Amog
  Kamsetty}, \bibinfo{person}{Brenton Chu}, {and} \bibinfo{person}{Nikita
  Vemuri}.} \bibinfo{year}{2018}\natexlab{}.
\newblock \showarticletitle{Targeted adversarial examples for black box audio
  systems}.
\newblock \bibinfo{journal}{\emph{arXiv preprint arXiv:1805.07820}}
  (\bibinfo{year}{2018}).
\newblock


\bibitem[\protect\citeauthoryear{Tram{\`e}r, Zhang, Juels, Reiter, and
  Ristenpart}{Tram{\`e}r et~al\mbox{.}}{2016}]%
        {tramer2016stealing}
\bibfield{author}{\bibinfo{person}{Florian Tram{\`e}r}, \bibinfo{person}{Fan
  Zhang}, \bibinfo{person}{Ari Juels}, \bibinfo{person}{Michael~K. Reiter},
  {and} \bibinfo{person}{Thomas Ristenpart}.} \bibinfo{year}{2016}\natexlab{}.
\newblock \showarticletitle{Stealing Machine Learning Models via Prediction
  {API}s}. In \bibinfo{booktitle}{\emph{USENIX Security Symposium}}.
  \bibinfo{publisher}{USENIX}, \bibinfo{pages}{601--618}.
\newblock


\bibitem[\protect\citeauthoryear{Voran and Sholl}{Voran and Sholl}{1995}]%
        {voran-1995-perception}
\bibfield{author}{\bibinfo{person}{Stephen Voran} {and} \bibinfo{person}{Connie
  Sholl}.} \bibinfo{year}{1995}\natexlab{}.
\newblock \showarticletitle{Perception-based Objective Estimators of Speech}.
  In \bibinfo{booktitle}{\emph{IEEE Workshop on Speech Coding for
  Telecommunications}}. IEEE, \bibinfo{pages}{13--14}.
\newblock


\bibitem[\protect\citeauthoryear{Wang and Gong}{Wang and Gong}{2018}]%
        {wang-18-hyperparameters}
\bibfield{author}{\bibinfo{person}{Binghui Wang} {and}
  \bibinfo{person}{Neil~Zhenqiang Gong}.} \bibinfo{year}{2018}\natexlab{}.
\newblock \showarticletitle{Stealing Hyperparameters in Machine Learning}. In
  \bibinfo{booktitle}{\emph{Symposium on Security and Privacy}}.
  \bibinfo{publisher}{IEEE}.
\newblock


\bibitem[\protect\citeauthoryear{Yakura and Sakuma}{Yakura and Sakuma}{2019}]%
        {yakura-2019-robust}
\bibfield{author}{\bibinfo{person}{Hiromu Yakura} {and} \bibinfo{person}{Jun
  Sakuma}.} \bibinfo{year}{2019}\natexlab{}.
\newblock \showarticletitle{Robust audio adversarial example for a physical
  attack}.
\newblock \bibinfo{journal}{\emph{arXiv preprint arXiv:1810.11793}}
  (\bibinfo{year}{2019}).
\newblock


\bibitem[\protect\citeauthoryear{Yang}{Yang}{1999}]%
        {yang-1999-enhancedmb}
\bibfield{author}{\bibinfo{person}{Wonho Yang}.}
  \bibinfo{year}{1999}\natexlab{}.
\newblock \emph{\bibinfo{title}{Enhanced Modified Bark Spectral Distortion
  (EMBSD): an Objective Speech Quality Measrure Based on Audible Distortion and
  Cognition Model}}.
\newblock \bibinfo{thesistype}{Ph.D. Dissertation}. \bibinfo{school}{Temple
  University Graduate Board}.
\newblock


\bibitem[\protect\citeauthoryear{Yuan, Chen, Zhao, Long, Liu, Chen, Zhang,
  Huang, Wang, and Gunter}{Yuan et~al\mbox{.}}{2018}]%
        {yuan-18-commandersong}
\bibfield{author}{\bibinfo{person}{Xuejing Yuan}, \bibinfo{person}{Yuxuan
  Chen}, \bibinfo{person}{Yue Zhao}, \bibinfo{person}{Yunhui Long},
  \bibinfo{person}{Xiaokang Liu}, \bibinfo{person}{Kai Chen},
  \bibinfo{person}{Shengzhi Zhang}, \bibinfo{person}{Heqing Huang},
  \bibinfo{person}{Xiaofeng Wang}, {and} \bibinfo{person}{Carl~A. Gunter}.}
  \bibinfo{year}{2018}\natexlab{}.
\newblock \showarticletitle{{CommanderSong}: A Systematic Approach for
  Practical Adversarial Voice Recognition}.
\newblock \bibinfo{journal}{\emph{arXiv preprint arXiv:1801.08535}}
  (\bibinfo{year}{2018}).
\newblock


\bibitem[\protect\citeauthoryear{Zhang, Yan, Ji, Zhang, Zhang, and Xu}{Zhang
  et~al\mbox{.}}{2017}]%
        {zhang2017dolphinattack}
\bibfield{author}{\bibinfo{person}{Guoming Zhang}, \bibinfo{person}{Chen Yan},
  \bibinfo{person}{Xiaoyu Ji}, \bibinfo{person}{Tianchen Zhang},
  \bibinfo{person}{Taimin Zhang}, {and} \bibinfo{person}{Wenyuan Xu}.}
  \bibinfo{year}{2017}\natexlab{}.
\newblock \showarticletitle{{DolphinAttack}: {I}naudible Voice Commands}. In
  \bibinfo{booktitle}{\emph{Conference on Computer and Communications Security
  (CCS)}}. \bibinfo{publisher}{ACM}, \bibinfo{pages}{103--117}.
\newblock


\bibitem[\protect\citeauthoryear{Zwicker and Fastl}{Zwicker and Fastl}{2007}]%
        {zwicker2013psychoacoustics}
\bibfield{author}{\bibinfo{person}{Eberhard Zwicker} {and}
  \bibinfo{person}{Hugo Fastl}.} \bibinfo{year}{2007}\natexlab{}.
\newblock \bibinfo{booktitle}{\emph{{Psychoacoustics}: {F}acts and Models}
  (\bibinfo{edition}{third} ed.)}.
\newblock \bibinfo{publisher}{Springer}.
\newblock


\end{thebibliography}

\onecolumn

\appendix
\section{Room Layout Plans}
\label{app:rooms}

\begin{table}[h]
\vspace{2em}
\caption{Microphone and Speaker positions and the reverberation time for each room in Table~\ref{tab:rooms}.}
\smallskip
\centering
\normalsize
\resizebox{0.8\columnwidth}{!}{
\begin{tabular}{l|c|c|c|c} 
\toprule
   & \multirow{2}{*}{$T_{60}$} & \multirow{2}{*}{Microphone} & \multicolumn{2}{c}{Speaker}  \\ [2pt] 
     %\cmidrule{4-5}
  & & & w/ line-of-sight & w/o line-of-sight  \\ [2pt] 
\midrule 
Lecture & \multirow{2}{*}{$0.80$\,s} & \multirow{2}{*}{$\mathbf{r}=[8.1\text{\,m}, 3.4\text{\,m}, 1.2\text{\,m}]$} & \multirow{2}{*}{$\mathbf{s}=[11.0\text{\,m}, 3.4\text{\,m}, 1.2\text{\,m}]$} & \multirow{2}{*}{$\mathbf{s}=[8.9\text{\,m}, 2.2\text{\,m}, 0.0\text{\,m}]$}  \\ 
Room  & & & &  \\ 
 \midrule 
Meeting & \multirow{2}{*}{$0.74$\,s} & \multirow{2}{*}{$\mathbf{r}=[3.7\text{\,m}, 5.7\text{\,m}, 1.2\text{\,m}]$} & \multirow{2}{*}{$\mathbf{s}=[1.8\text{\,m}, 5.7\text{\,m}, 1.2\text{\,m}]$} & \multirow{2}{*}{$\mathbf{s}=[3.7\text{\,m}, 4.9\text{\,m}, 0.0\text{\,m}]$} \\ 
Room  & & & &  \\ 
 \midrule 
%Small & \multirow{2}{*}{$0.50$\,s} & \multirow{2}{*}{$\mathbf{r}=[0.2\text{\,m}, 2.6\text{\,m}, 1.2\text{\,m}]$} & \multirow{2}{*}{$\mathbf{s}=[1.7\text{\,m}, 0.6\text{\,m}, 1.2\text{\,m}]$} & \multirow{2}{*}{$\mathbf{s}=[1.7\text{\,m}, 2.4\text{\,m}, 0.0\text{\,m}]$} & \multirow{2}{*}{$\mathbf{s}=[0.2\text{\,m}, 0.8\text{\,m}, 1.2\text{\,m}]$} \\ 
%Office & & & & & \\ 
% \midrule 
\multirow{2}{*}{Office} & \multirow{2}{*}{$0.64$\,s} & \multirow{2}{*}{$\mathbf{r}=[3.8\text{\,m}, 1.8\text{\,m}, 1.2\text{\,m}]$} & \multirow{2}{*}{$\mathbf{s}=[1.4\text{\,m}, 4.6\text{\,m}, 1.2\text{\,m}]$} & \multirow{2}{*}{$\mathbf{s}=[-0.5\text{\,m}, 2.0\text{\,m}, 1.2\text{\,m}]$} \\ 
 & & & &  \\ 
\bottomrule
\end{tabular}
}
\label{tab:positions}
\end{table}

% LECTURE ROOM
\begin{figure}[!htbp]

\begin{minipage}{1.0\textwidth}
\begin{flushright}
\includegraphics[width=17.13cm, keepaspectratio]{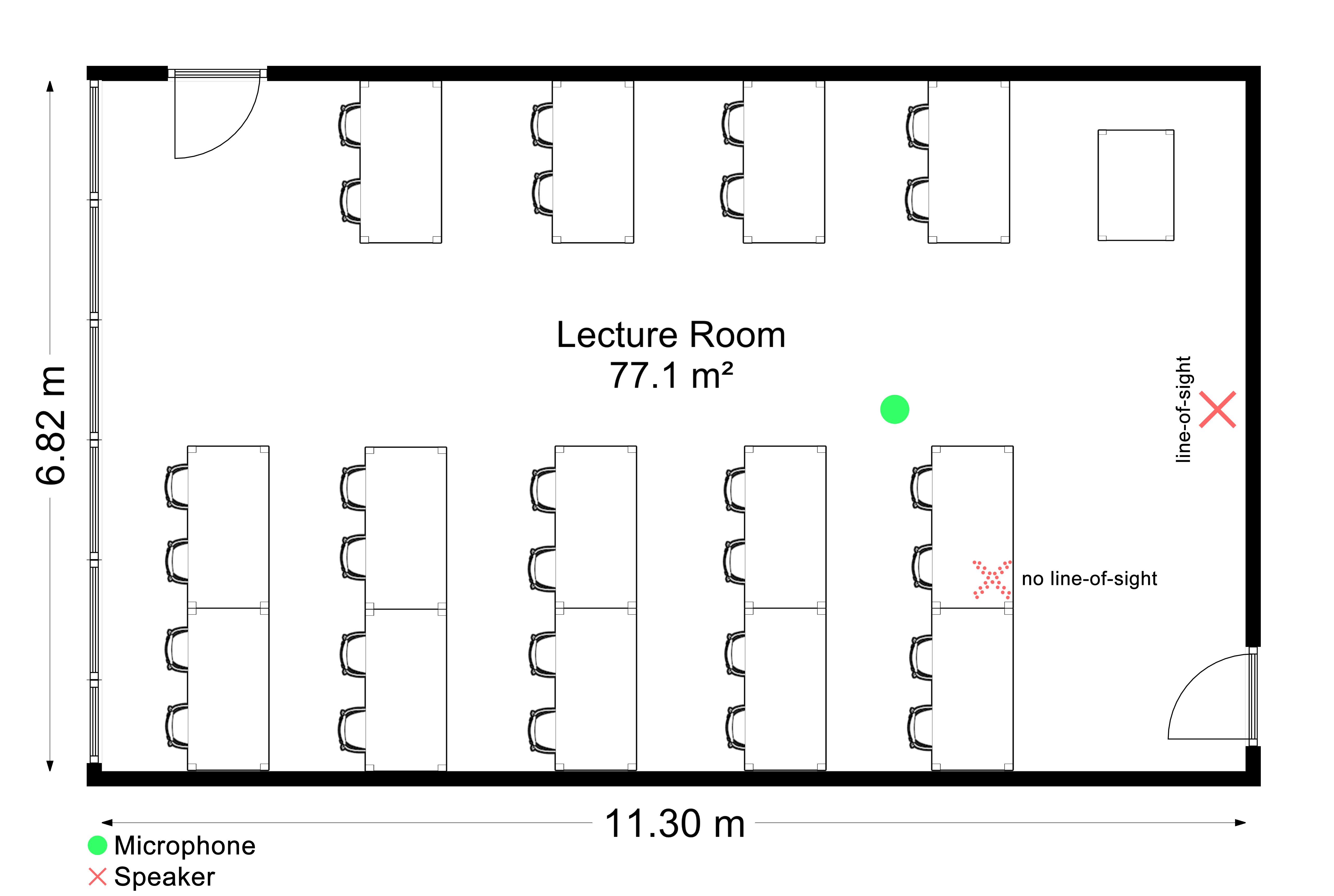}
\caption{Room layout of the lecture room.}
\label{fig:lecture}
\end{flushright}
\end{minipage}
\end{figure}
% LARGE OFFICE
\begin{figure}[!htbp]
\vspace{2em}
\begin{minipage}{1.0\textwidth}
\centering
%\begin{flushright}
\includegraphics[width=13cm, keepaspectratio]{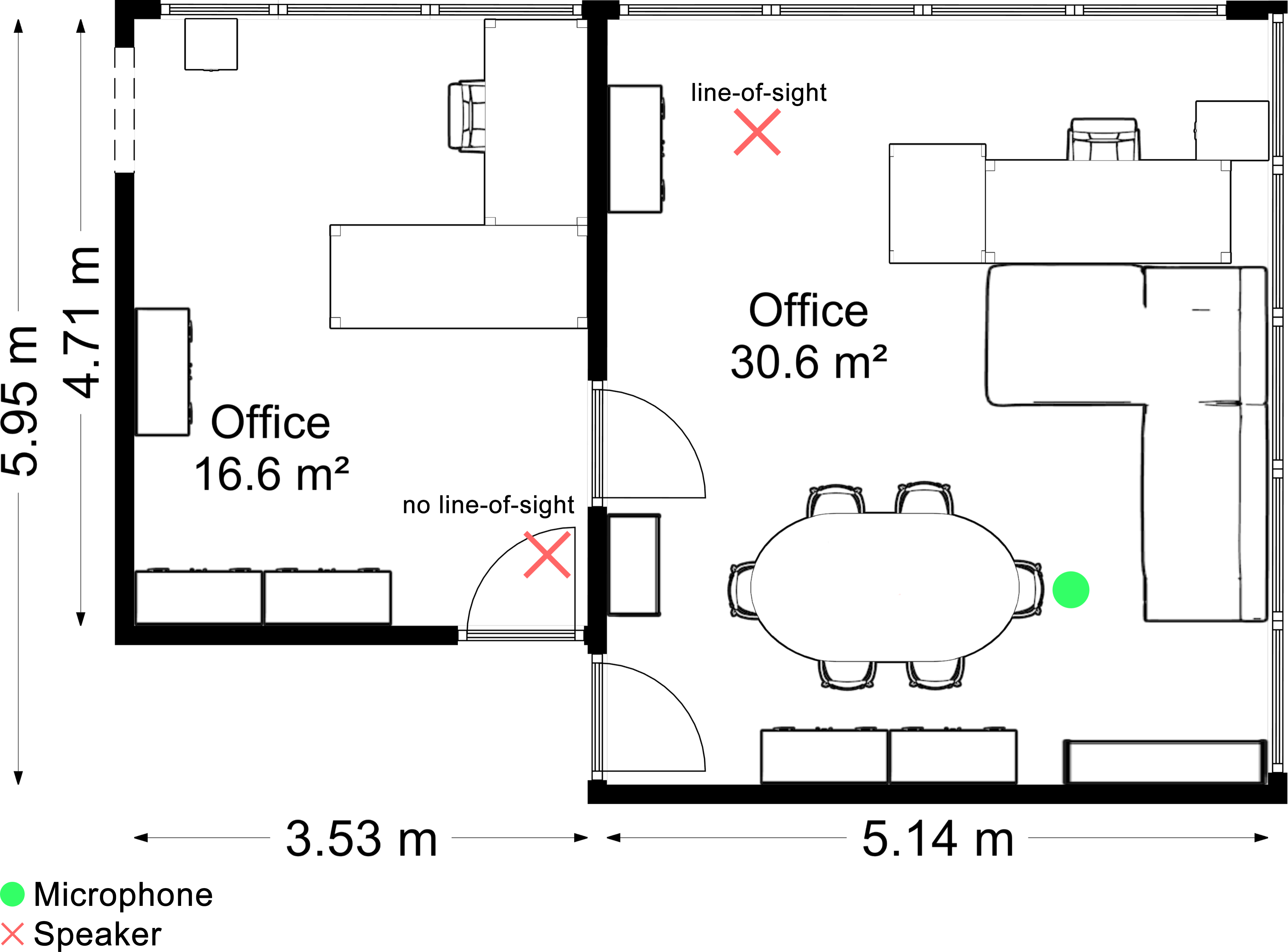}
\caption{Room layout of the office room.}
\label{fig:largeoffice}
%\end{flushright}
\end{minipage}
\end{figure}

%  MEETING ROOM
\begin{figure}[!htbp]
\vspace{2em}
\begin{minipage}{1.0\textwidth}
\centering
%\begin{flushright}
\includegraphics[width=9cm, keepaspectratio, angle=-90]{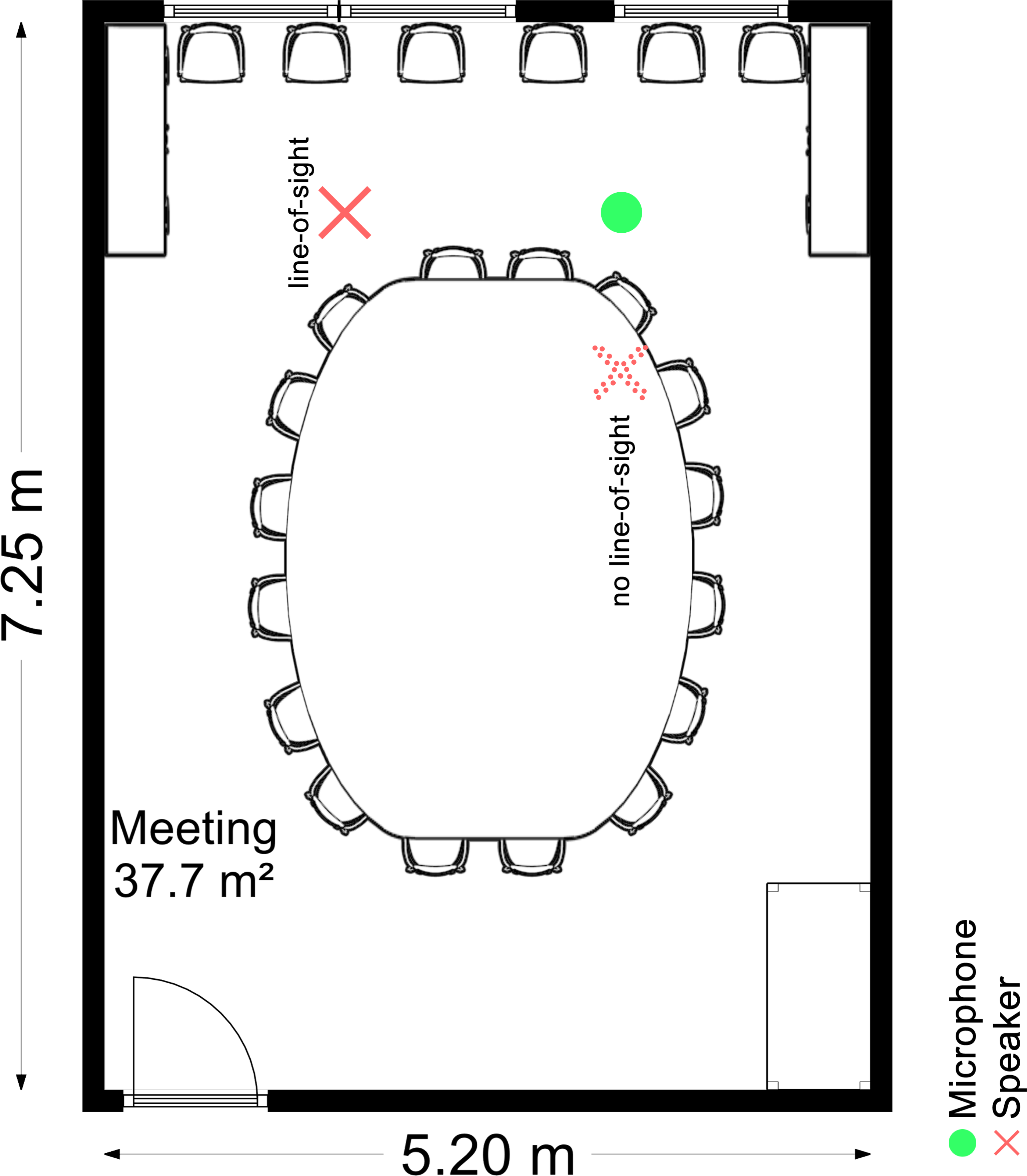}
\caption{Room layout of the meeting room.}
\label{fig:meetingoffice}
%\end{flushright}
\end{minipage}
\end{figure}

% SMALL OFFICE AND MEETING ROOM
%\begin{figure}[!htbp]
%\begin{minipage}{0.436077\textwidth}
%\begin{flushright}
%\includegraphics[width=7.47cm, keepaspectratio]{figs/1_Small.png}
%\caption{Room layout of the small office room.}
%\label{fig:smalloffice}
%\end{flushright}
%\end{minipage}
%\begin{minipage}{0.563923\textwidth}
%\begin{flushright}
%\includegraphics[width=9.66cm, keepaspectratio]{figs/1_Meeting.png}
%\caption{Room layout of the meeting room.}
%\label{fig:meetingroom}
%\end{flushright}
%\end{minipage}
%\end{figure}

\end{document}